\documentclass[aps,prl,twocolumn,showpacs,superscriptaddress,groupedaddress]{revtex4-1}  % for double-spaced preprint
\usepackage{graphicx}  % needed for figures
\usepackage{dcolumn}   % needed for some tables
\usepackage{bm}        % for math
\usepackage{amssymb}   % for math
\usepackage{amsmath}
\usepackage{epstopdf}
\usepackage{siunitx}

\begin{document}

% The following information is for internal review, please remove them for submission
\widetext
%\leftline{Version xx as of \today}
%\leftline{Primary authors: Joe E. Physics}
%\leftline{To be submitted to (PRL, PRD-RC, PRD, PLB; choose one.)}
%\leftline{Comment to {\tt d0-run2eb-nnn@fnal.gov} by xxx, yyy}
%\centerline{\em D\O\ INTERNAL DOCUMENT -- NOT FOR PUBLIC DISTRIBUTION}

% the following line is for submission, including submission to the arXiv!!
%\hspace{5.2in} \mbox{Fermilab-Pub-04/xxx-E}

%\title{Nonlinear traveling waves with oscillatory tails in single column woodpile chains}
\title{Global Bifurcations in a Damped-Driven Diatomic Granular Crystal}
%\title{Periodic Wave Propagation Inside a Highly Nonlinear Acoustic
%Metamaterial -- More Provocatively Perhaps: ``Is the Sonic Vacuum Really a Vacuum ?'' }
%\input author_list.tex       % D0 authors (remove the first 3 lines
                             % of this file prior to submission, they
                             % contain a time stamp for the authorlist)
                             % (includes institutions and visitors)
\author{D. Pozharskiy}
\affiliation{Department of Chemical and Biological Engineering, Princeton University, Princeton, NJ 08544, USA}

\author{Ioannis G. Kevrekidis}
\affiliation{Departments of Chemical and Biomolecular Engineering \& of Applied Mathematics and Statistics, Johns Hopkins University, Baltimore, MD 21218, USA}

\author{Panayotis G. Kevrekidis}
\affiliation{Department of Mathematics and Statistics, University of Massachusetts, Amherst, 01003-4515, Massachusetts, USA}

\date{\today}

\begin{abstract}
We revisit here the dynamics of an
engineered dimer granular crystal under an external periodic drive
in the presence of dissipation. Earlier findings included a saddle-node bifurcation, whose terminal point initiated the observation of chaos; the system was found to 
exhibit bistability and potential quasiperiodicity.
We now complement these findings by the identification of
unstable manifolds of saddle periodic solutions (saddle points of the stroboscopic map) within the system 
dynamics. We unravel how homoclinic tangles of these manifolds
lead to the appearance of a chaotic attractor,
upon the apparent period-doubling bifurcations that destroy 
invariant tori associated with quasiperiodicity.
These findings complement the earlier ones, offering
more concrete insights into the emergence of chaos within this high-dimensional, experimentally accessible system.
\end{abstract}

%\pacs{45.70.-n 05.45.-a 46.40.Cd}
\maketitle

%{\bf Introduction.} 
\section{Introduction}

The realm of engineered granular crystals has
proven over the last few decades to be an excellent
testbed for a remarkable breadth of ideas at the
nexus of nonlinear science and mechanical 
metamaterials~\cite{nesterenko2001dynamics,sen2008solitary,vakakis_review,yuli_book,granularBook}
The associated closely packed arrays of particles interact through  nonlinear Hertzian contacts~\cite{nesterenko2001dynamics,sen2008solitary}.
%their dynamics exhibit a wide range of phenomena due to their nonlinearities, which include solitary waves
Their versatility, ranging from homogeneous
to markedly heterogeneous systems, but more importantly
ranging from weakly to strongly nonlinear regimes,
has enabled the probing of a wide range of coherent
structures.
These include, but are not limited to, the propagation of traveling waves~\cite{nesterenko2001dynamics,sen2008solitary,nesterenko1983propagation,lazaridi1985observation,coste1997solitary,daraio2006tunability,porter2008highly,starosvetsky2010traveling,jayaprakash2011new}, the instabilities leading to the formation of
discrete breathers both in a bright~\cite{boechler2010discrete,theocharis2010intrinsic}, and in a dark form~\cite{chong2014damped}, the disintegration of shocks~\cite{daraio2006energy} and their formation~\cite{herbold2007shock,molinari2009stationary,KdVToda_limits} (see also the related setting of~\cite{talcohen})
and
nonlinear resonances~\cite{pozharskiy2015nonlinear,zhang2017experimental}.
This flexibility have made them a potential candidate platform for many engineering applications including shock and energy-absorbing layers~\cite{daraio2006energy,fraternali2009optimal},
acoustic lenses~\cite{spadoni2010generation}, actuating devices~\cite{khatri2008highly}, sound scramblers~\cite{nesterenko2005anomalous,daraio2005strongly},
acoustic switches~\cite{li2014granular} and rectifiers~\cite{boechler2011bifurcation}.

Although many studies have focused on Hamiltonian systems, dissipation in granular crystals is important and has also been a subject of intense investigations~\cite{rosas2007observation,rosas2008short,carretero2009dissipative,vergara2010model}, some of which
are still ongoing~\cite{James_2021}.
In addition, adding external forcing to a dissipative granular crystal, we can observe periodic solutions in the system resulting from the relevant frequency interplay~\cite{boechler2011bifurcation,hoogeboom2013hysteresis,chong2014damped,charalampidis2015time,lydon2015nonlinear,pozharskiy2015nonlinear}; see also~\cite{pdimer} for a 
recent example.
Better understanding of how signals of varying amplitude and frequency propagate through the granular system, can be achieved by performing a systematic exploration of the parameter space using numerical continuation of solutions~\cite{doedel1991numerical}.
Local bifurcations can be located by inspecting the linearization around the solutions, unveiling additional characteristics of the system such as loss of stability, secondary solution branches, multiple attractors or quasiperiodicity~\cite{guckenheimer1983nonlinear}.
However, in certain cases the dynamical behavior of the system changes dramatically and this seems to happen unexpectedly, as the linearization is not sufficient to pinpoint and explain these changes. This phenomenon is due to {\em global bifurcations}, which involve the interactions
of stable and unstable manifolds of steady states or periodic solutions in the system; these invariant manifold interactions can lead to the creation or destruction of attractors, as well as give rise to chaotic behavior in the system~\cite{guckenheimer1983nonlinear,kevrekidis1987numerical}.

A previous study of some of the present authors concerned a one-dimensional, statically compressed, diatomic granular crystal with driving frequency inside the band gap of the linear spectrum~\cite{hoogeboom2013hysteresis}. There, it was determined that as
the amplitude of the forcing increases, two time-periodic solutions disappear in a saddle-node bifurcation and afterwards the response of the system would ``jump''
to a chaotic branch, which constitutes the stable attractor of
the dynamics for higher parameter values. This switch to chaotic behavior is accompanied by increased energy
propagation through the granular chain. Additionally, the system was found to exhibit hysteresis where periodic and chaotic attractors coexist as well as potential regions of quasiperiodicity.

Building upon the previous results, our goal in this work was to suggest a potential explanation for the emergence of chaos in the system by studying the one-dimensional (in the stroboscopic map; actually two-dimensional), unstable manifolds of saddle points (in the stroboscopic map; actually, saddle-type periodic solutions).
We will show how homoclinic tangles of the aforementioned manifolds are involved in the appearance of the chaotic attractor. Additionally, we will show how invariant tori associated with quasiperiodicity are destroyed
in a sequence of (apparent) period-doubling bifurcations. Using optimization techniques we were also able to numerically converge to a new branch of solutions, not visibly connected to the already known branches. One-dimensional, unstable manifolds of saddles
on the new branch are shown to also be involved in the chaotic dynamics and may be the reason for the hysteretic dynamics in the system. Our results provide a description of the global bifurcation structure
in the system, as comprehensive as possible given the complex and high-dimensional nature of the system dynamics.

%{\bf Configuration and model.} 
\section{Theoretical Setup}

The system consists of a precompressed, diatomic granular crystal with $N = 20$ alternating aluminum and steel
spheres, where the first sphere is an aluminum one. The aluminum spheres have radius $R_a = 9.53$ mm, mass $m_a = 9.75$ g, elastic modulus
$E_a = 73.5$ GPa, Poisson ratio $\nu_a = 0.33$ and the steel spheres have radius $R_b = R_a$, mass $m_b = 28.84$ g, elastic modulus
$E_b = 193$ GPa and Poisson ratio $\nu_b = 0.3$. The spheres are constrained in a one-dimensional configuration and on one side
there is a piezoelectric actuator that drives the system at a specified frequency. The other side is attached to a soft spring (stiffness $1.24$ \si{kN/m}),
that is used to precompress the system with a static load ($F_0 = 8$ N). Indeed, here, we are reiterating the
configuration experimentally set up in~~\cite{hoogeboom2013hysteresis}.

The system is modeled in that work as a damped-driven granular crystal and can be described by the following system of equations:
\begin{eqnarray} \label{eq:diatomicGranular}
 &m_j\acute{\ddot{u}}_j = A_{j-1,j}\left[\delta_{j-1,j}-\left(\acute{u}_j-\acute{u}_{j-1}\right)\right]_{+}^{3/2} \\ \nonumber
 &-A_{j,j+1}\left[\delta_{j,j+1}-\left(\acute{u}_{j+1}-\acute{u}_j\right)\right]_{+}^{3/2} - \frac{m_j\acute{\dot{u}}_j}{\tau}, 
\end{eqnarray}
where $m_j$ is the mass of the $j$-th sphere, $A_{j,j+1} = \frac{4E_jE_{j+1}\sqrt{\frac{R_jR_{j+1}}{R_j+R_{j+1}}}}{3E_{j+1}(1-\nu_j^2)+3E_j(1-\nu_{j+1}^2)}$
is the coefficient of the Hertzian contact and $\delta_{j,j+1} = (F_0/A_{j,j+1})^{2/3}$ is the static overlap between spheres $j$ and $j+1$ due to the precompression.
The actuator provides an external, periodic forcing on the system at its left end and can be well approximated
as a $0$-th sphere with displacement $u_0(t) = \alpha\cos{(2\pi f_dt)}$, where
$\alpha$ is the amplitude and $f_d$ the frequency of the periodic motion. The brackets $[\cdot]_+ = \max{(\cdot,0)}$ are used to indicate that there is a force between
spheres only when they are in contact, while there
is no force when the spheres are separated (i.e., when the
argument is negative). The dissipation is considered to be linear and on site with time constant $\tau = 1.75$ ms~\cite{boechler2011bifurcation}.
Using $\bar{m} = \frac{m_a}{m_b}$, $\bar{t} = \sqrt{\frac{\bar{m}}{A_{j,j+1}\sqrt{\delta_{j,j+1}}}}$ and $u_j = \frac{\acute{u}_j}{\delta_{j,j+1}}$ as
the nondimensional position, the equation (\ref{eq:diatomicGranular}) becomes:
\begin{eqnarray}
 &\ddot{u}_j = \frac{\bar{m}}{m_j}\left[1-\left(u_j-u_{j-1}\right)\right]_{+}^{3/2} \\ \nonumber
 &-\frac{\bar{m}}{m_j}\left[1-\left(u_{j+1}-u_j\right)\right]_{+}^{3/2} - \frac{\dot{u}_j}{\acute{\tau}},
\end{eqnarray}
where $\acute{\tau}$ is a scaled dissipation time constant. Additionally, we define $v_j = \dot{u}_j$ as the velocity of the $j$-th sphere.
While we recognize that the precise form of the dissipation
characterizing the experiment remains a topic of active
investigation~\cite{rosas2007observation,rosas2008short,carretero2009dissipative,vergara2010model,James_2021}, the 
perspective adopted herein is similar to that
of~\cite{hoogeboom2013hysteresis} whereby this simpler
dash-pot approximation is sufficient to capture the essence of
the experimental findings.
The frequency $f_d = 6$ kHz which was used for all subsequent results lies inside the band gap for the given precompressive force~\cite{hoogeboom2013hysteresis}.

%{\bf Numerical tools.} 

\section{Numerical Methods}

A widely used tool to study periodic dynamical systems is the Poincar\'{e} map. Instead of observing the trajectory of the dynamical
system continuously in time, we choose a hyperplane $\Pi$ in phase space and observe the system state only when the trajectory crosses $\Pi$. A particularly
convenient version of the Poincar\'{e} map for forced systems is the {\em stroboscopic map}, where the system state is observed once every period of the forcing $T$.
Entrained periodic solutions have the same period as (resp. integer multiples of the period of) the forcing and appear as fixed points (resp. periodic points) of the stroboscopic map. Quasiperiodic solutions (two incommensurate frequencies)
appear as closed curves (invariant circles) of the map.

The fixed (resp. periodic) points of the map at a given amplitude of the forcing were found using Newton's method and
pseudoarclength continuation was used to numerically track the different branches of solutions as the forcing amplitude changes~\cite{doedel1991numerical}.
Additionally, the eigenvalues of the linearization around a fixed point provide stability information, as they are the Floquet multipliers (FMs) of the respective limit cycle.
A fixed point is stable if all of the FMs are inside the unit circle. Fixed points with real eigenvalues that have modulus greater than one are called {\em saddles} and
constitute the main object of study in our work. Saddle points have stable and unstable invariant manifolds, $W^s$ and $W^u$ respectively. Their interactions cause global
bifurcations in the system, hence the computation of these manifolds is the main tool we use to locate the parameter regions where these bifurcations occur and what
change in the dynamical behavior of the system they bring. Of particular interest are saddles with only a single unstable eigenvalue, since in this case the (stroboscopic) unstable manifold
$W^u$ is one-dimensional and easy to compute. In order to do so, we select a point in the direction of the unstable eigenvector $E^u$, which is tangent to $W^u$ in the
vicinity of the saddle~\cite{guckenheimer1983nonlinear}, and iterate this point using the stroboscopic map. We improve the resolution of the manifold by interpolating
points between two successive iterates of the map and iterating them as well~\cite{hobson1993efficient}. Computation of two-dimensional manifolds is also possible~\cite{krauskopf1998growing},
however we did not perform it in the present work (meaningful visualization of complex interactions of 2D manifolds in high-dimensional spaces is humanly impractical).

%{\bf Results.} 
\section{Computational Findings}

\begin{figure}[htbp]
\centering
\includegraphics[width=\linewidth]{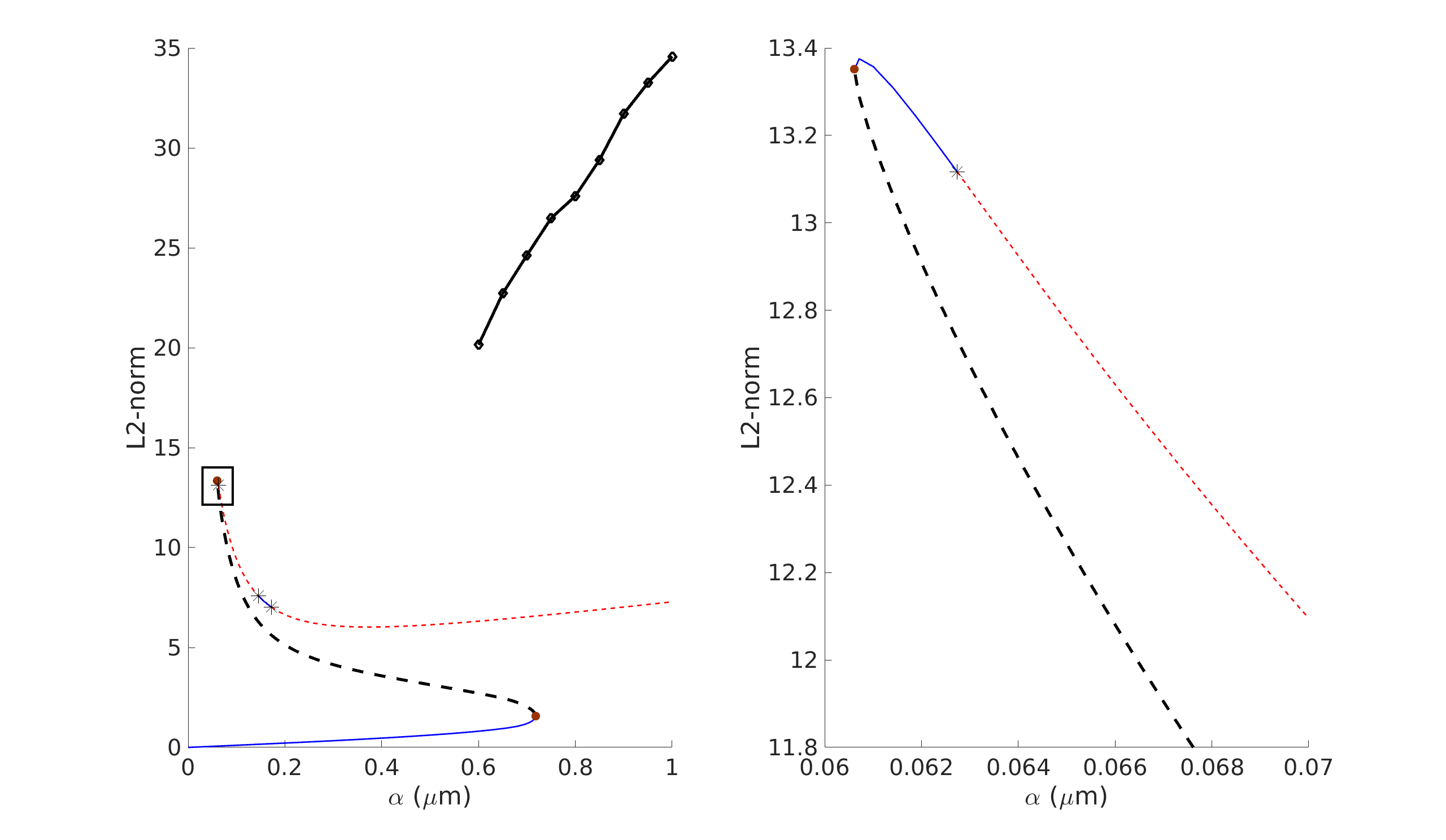}
\caption{Original bifurcation diagram from~\cite{hoogeboom2013hysteresis} and a zoom-in of the top peak. Stable branches are solid blue, unstable branches are dashed red, saddle-node bifurcations are shown as red dots and Neimark-Sacker bifurcations as black stars.
The branch of saddle points with one-dimensional, unstable manifolds is colored black for easier reference. The high-amplitude connected black dots denote the chaotic attractor, see text.}
\label{fig:origBif}
\end{figure}

We begin the discussion of our results by briefly reiterating the original bifurcation diagram obtained in~\cite{hoogeboom2013hysteresis}. Fig.~\ref{fig:origBif} shows the norm
of every individual periodic solution obtained at different amplitudes of the forcing; notice that the bifurcation
diagram features the same structural characteristics even
though a different bifurcation diagnostic was used 
in~\cite{hoogeboom2013hysteresis} in order to connect
with physical experiments. Stable branches are shown as solid lines and unstable branches as dashed lines.
As we increase the amplitude from zero the system has a stable solution corresponding to the lowest (entrained, ``period-one") branch. The norm of the solution is low since energy doesn't propagate through
the granular chain as the forcing frequency is inside the band gap. 
This branch can, structurally, be thought of being ``slaved''
to the external drive.
Eventually this branch is lost in the saddle node bifurcation (red dots) at an amplitude $\alpha_{sn}^1 \approx 0.71$ \si{\mu {\rm m}}.

After the saddle-node bifurcation, no stable 
(periodic orbit) attractor exists and the behavior of the system is chaotic.
Fig.~\ref{fig:origBif} also includes this chaotic branch (black line with diamonds), which was computed by integrating the system at specified amplitude
values for 1000 periods and averaging the norm of the state after each period. We begin the computation at the largest amplitude value and step backwards in small
increments since the chaotic branch exhibits hysteresis. It disappears completely at an amplitude $\alpha_{ch} \approx 0.59$ \si{\mu {\rm m}}, after which the system dynamically converges
to the lowest stable branch regardless of the initial condition. We reiterate here that the features of $\alpha_{sn}^1$ and 
$\alpha_{ch} $ were also reported in the original
study of~\cite{hoogeboom2013hysteresis}.

The unstable branch that was born through the saddle-node bifurcation has a single unstable eigenvalue, hence we can compute its (stroboscopically) one-dimensional unstable manifold; this will uncover the first global
bifurcation in the system as we will see later on. Tracing this branch backwards, it disappears again through a saddle-node bifurcation at amplitude $\alpha_{sn}^2 \approx 0.06$ \si{\mu {\rm m}}.
A new, stable for a small parametric interval branch, results (better seen in Fig.~\ref{fig:origBif}b). 
It is important to appreciate that, as the dissipation 
coefficient decreases, this branch will be born
progressively closer to the $\alpha \rightarrow 0$ limit, i.e., this branch pertains to the ``bulk dimer breather''
which is intrinsic to the Hamiltonian version of the
dynamical system of interest. Once again, this is
in contra-distinction to the small amplitude branch
effectively dictated by the driver.
This upper, third branch loses stability through a supercritical Neimark-Sacker (NS) bifurcation (black stars) and a stable $T^2$ torus appears. 
At this
point two stable attractors coexist - the torus (stroboscopic invariant circle) and the lowest stable branch - and a dynamical trajectory can converge to either depending on the initial state.
The basin boundary should correspond to the ``middle" saddle (unstable) branch and its high-dimensional (codmension one) stable manifold.

Further to the right of this saddle-node bifurcation, there exists a short stable segment on the upper branch which gains and loses stability through a pair of Neimark-Sacker (NS)  bifurcations. The bifurcations appear to be subcritical as no stable torus was found in their vicinity.

%{\it Torus bifurcations.}

\begin{figure*}[htbp]
\centering
\includegraphics[width=\linewidth]{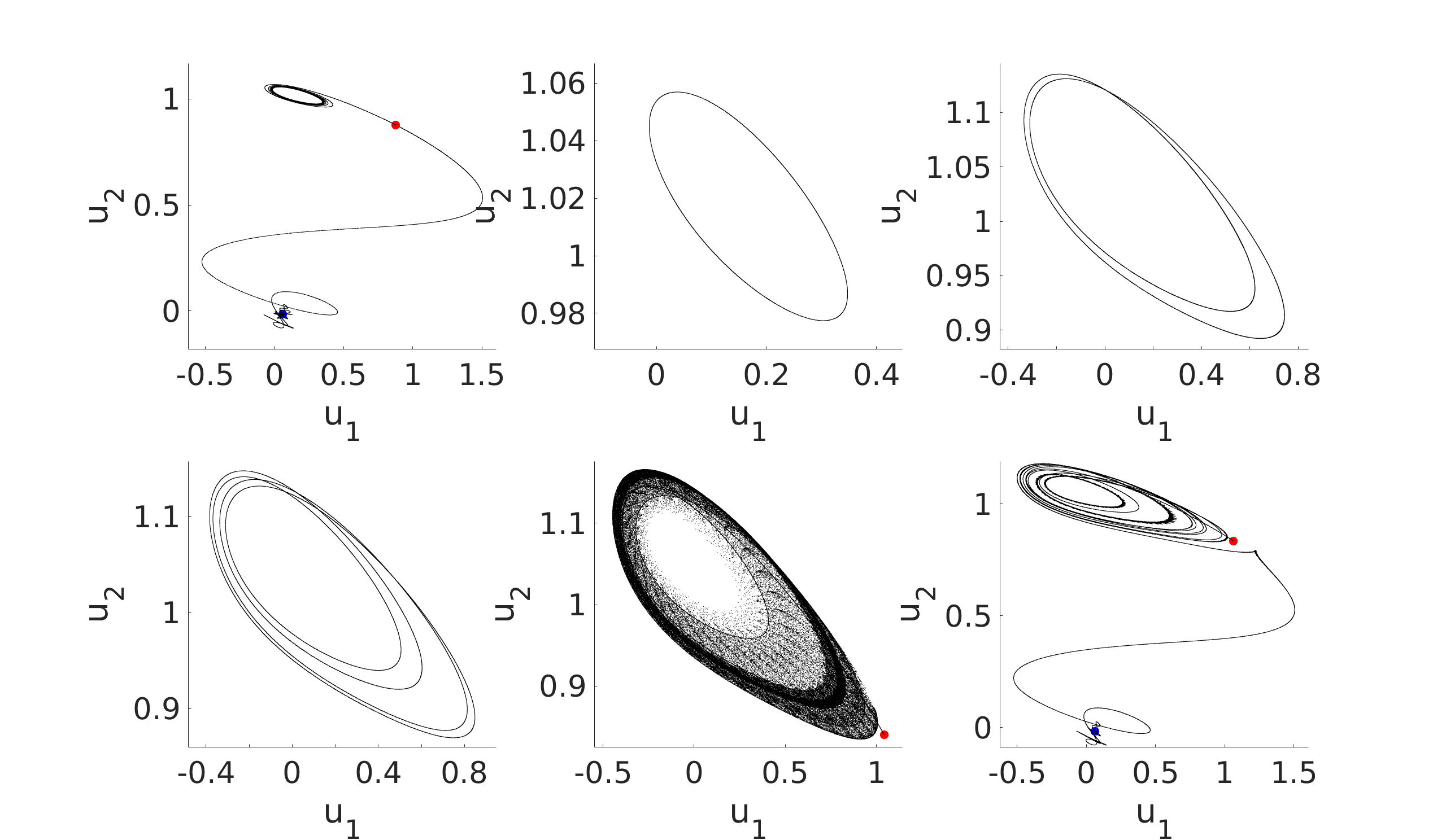}
\caption{Sequence of (apparent) period-doubling bifurcations of the torus and its eventual destruction. (a) $\alpha = 0.063$ $\mu$m, one side of the unstable manifold is attracted to the invariant circle and the other side to the low, stable branch (blue dot).
(b) $\alpha = 0.063$ $\mu$m, the invariant circle. (c) $\alpha = 0.065$ $\mu$m, first period-doubling bifurcation. (d) $\alpha = 0.06545$ $\mu$m, second period-doubling bifurcation. (e) $\alpha = 0.0661$ $\mu$m, no discernible invariant curve exists anymore.
(f) $\alpha = 0.0666$ $\mu$m, the side of the manifold previously attracted to the torus is eventually attracted to the stable branch as well.}
\label{fig:firstTorus}
\end{figure*}

\subsection{Torus Bifurcations}

Going back to the appearance of the torus, one side of the unstable manifold of the saddle is attracted to the invariant circle and the other to the lowest, stable branch. This can be seen in Fig.~\ref{fig:firstTorus}a, where the Poincar\'{e} map
is projected on the plane of position of the first and second bead. The saddle is shown as a red dot and the stable fixed point on the lowest branch as a blue dot. In the following panels the side that is attracted to the stable fixed point will be omitted.
Fig.~\ref{fig:firstTorus}b shows the final invariant circle. Increasing the amplitude further we observe that the invariant circle starts losing its smoothness and
the curve begins separating into two curves. After this transition the invariant curve is smooth again but now has a double loop structure (Fig.~\ref{fig:firstTorus}c). We say that the torus has undergone an {\em apparent} period-doubling bifurcation. This transition happens again and
the invariant curve now acquires a quadruple loop (Fig.~\ref{fig:firstTorus}d). Eventually (i.e., upon subsequent bifurcations), there is no discernible curve anymore to which the manifold is attracted to; however the manifold still wanders around the same region of the phase space where
the invariant curve existed. This can be seen in Fig.~\ref{fig:firstTorus}e, where the manifold covers a large region of the phase space without converging anywhere.
At higher amplitudes the side of the manifold that was attracted to this region, eventually ultimately converges to the lowest stable branch as well (Fig.~\ref{fig:firstTorus}f). At this point only one attractor exists.
The entire sequence of transitions and the final invariant curves can be seen in Fig.~\ref{fig:firstTorus}. This sequence of period-doubling bifurcations of a $T^2$ torus and its eventual destruction along with the appearance of a chaotic
regime has been qualitatively described in~\cite{kaneko1983doubling,franceschini1983bifurcations,iooss1988quasi}. We were able to identify clearly the first two apparent period-doubling bifurcations; however, it was increasingly harder
to identify any subsequent period-doublings before the destruction of the torus happens. We speculate that, after these period doublings, a very narrow (in parameter range) chaotic regime results, Fig.~\ref{fig:firstTorus}e, eventually destroyed in a global bifurcation (see below). 
%Additionally, no chaotic regime was identified and if it exists, it must be inside a very narrow parameter range.
%{\bf remark to be erased: do you mean past 2(f) here? Perhaps it 'd be
%good to be more specific or, if needed, a bit more speculative.}

%{\it Manifold tangles.}

\subsection{Saddle Invariant Manifold Tangles}

\begin{figure*}[htbp]
\centering
\includegraphics[width=\linewidth]{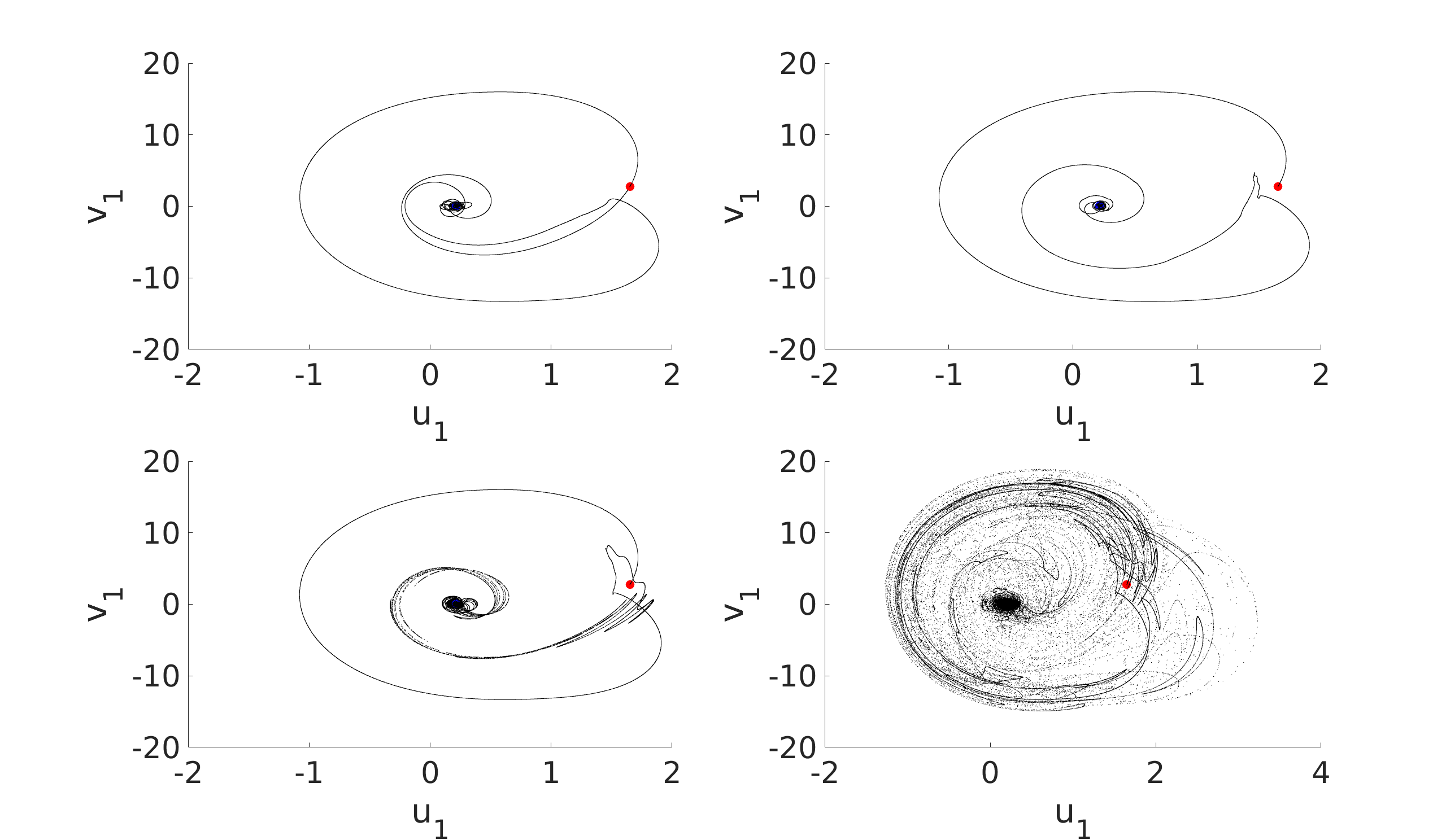}
\caption{First appearance of transient chaos. (a) $\alpha = 0.223$ $\mu$m, both sides of the unstable manifold appear ``smooth'' and are attracted to the low, stable branch. (b) $\alpha = 0.2245$ $\mu$m, one side of the manifold encounters another object and becomes ``squiggly''.
(c) $\alpha = 0.2249$ $\mu$m, the distortion of the manifold is more profound and resembles homoclinic tangencies. (d) $\alpha = 0.225$ $\mu$m, the manifold has become entangled and the system appears to be in transient chaos.}
\label{fig:blackSaddle1}
\end{figure*}

After the torus is destroyed, both sides of the unstable manifold of the saddle are attracted to the lowest stable branch. This picture persists for a wide parameter range as we move to the right on the bifurcation diagram. 
Fig.~\ref{fig:blackSaddle1}a shows the saddle and the two sides of its unstable manifold at an amplitude $\alpha = 0.223$ $\mu$m. Both sides appear to be ``smooth'' and quickly attracted to the stable fixed point. Just further to the right, one side of the manifold
encounters another object and the manifold appears ``squiggly'' for a small part of it (Fig.~\ref{fig:blackSaddle1}b). This phenomenon becomes more and more pronounced and it starts resembling homoclinic tangencies (Fig.~\ref{fig:blackSaddle1}c),
where there is an infinite number of tangencies between an unstable manifold and a stable manifold. These tangencies lead to manifold tangles that are known to be associated with chaotic behavior and this can be clearly seen in Fig.~\ref{fig:blackSaddle1}d, where the dynamics of the
system can be described as transient chaos, as one side of the manifold jumps erratically around the phase space. In all cases the other side of the unstable manifold is attracted to the lowest stable branch and has been omitted in the figures.

\begin{figure*}[htbp]
\centering
\includegraphics[width=\linewidth]{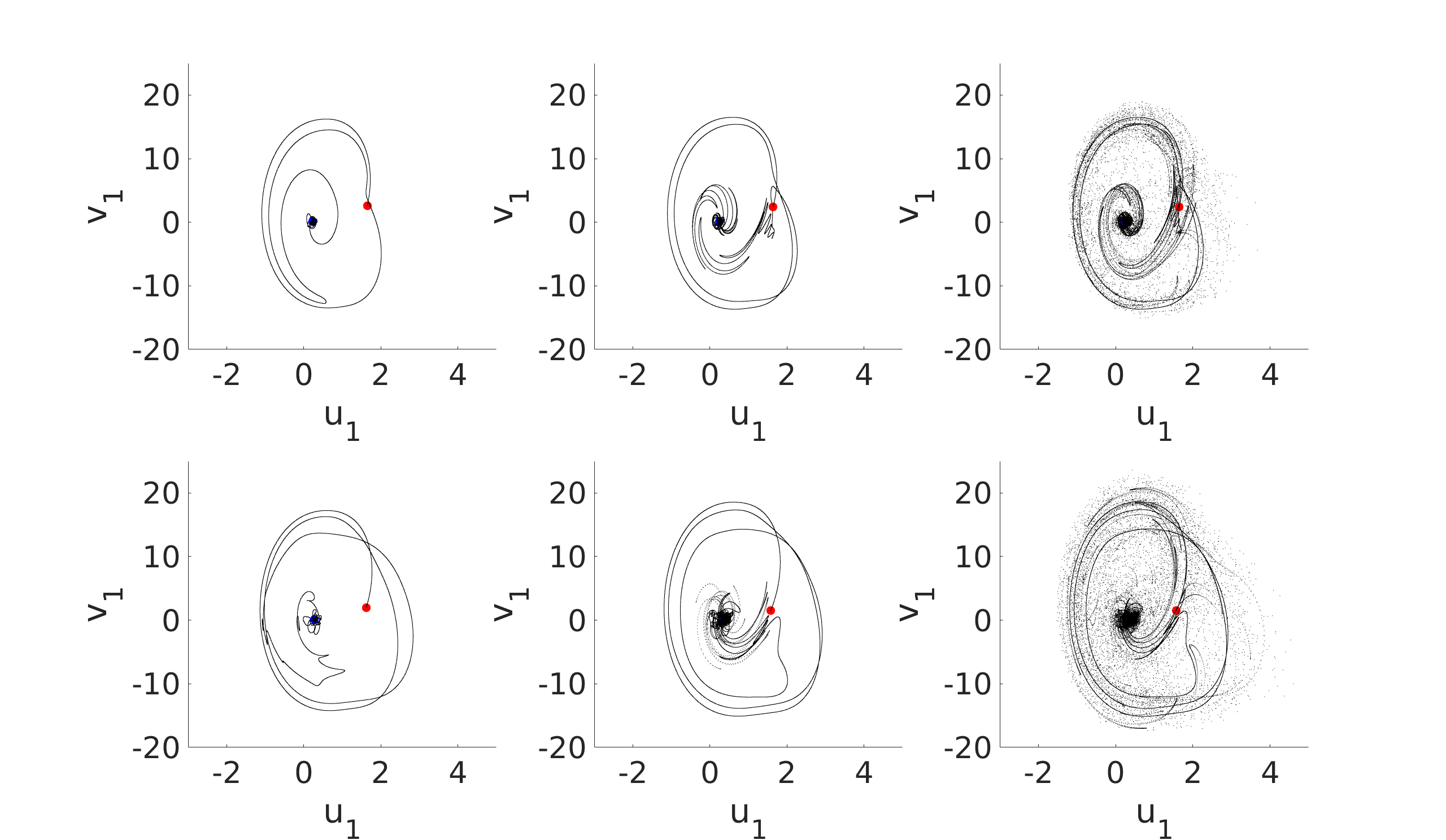}
\caption{The tangling and untangling of the black manifold happens multiple times. (a) $\alpha = 0.235$ $\mu$m. (b) $\alpha = 0.2494$ $\mu$m. (c) $\alpha = 0.2495$ $\mu$m. (d) $\alpha = 0.29$ $\mu$m. (e) $\alpha = 0.362$ $\mu$m. (f) $\alpha = 0.365$ $\mu$m.}
\label{fig:blackSaddle2}
\end{figure*}

This sequence of transitions happens several times as we increase the amplitude, where one side of the black unstable manifold becomes entangled and then untangles/disentangles as is seen in the different snapshots in Fig.~\ref{fig:blackSaddle2}. Eventually, the black manifold remains entangled
after an amplitude of $\alpha \approx 0.36$ \si{\mu {\rm m}}, until the saddle is lost in the saddle-node bifurcation. The volume that the entangled manifold visits in the phase space increases as we increase the amplitude of the forcing.

%{\it Extended bifurcation diagram.}

\subsection{Extended bifurcation diagram}

\begin{figure*}[htbp]
\centering
\includegraphics[width=\linewidth]{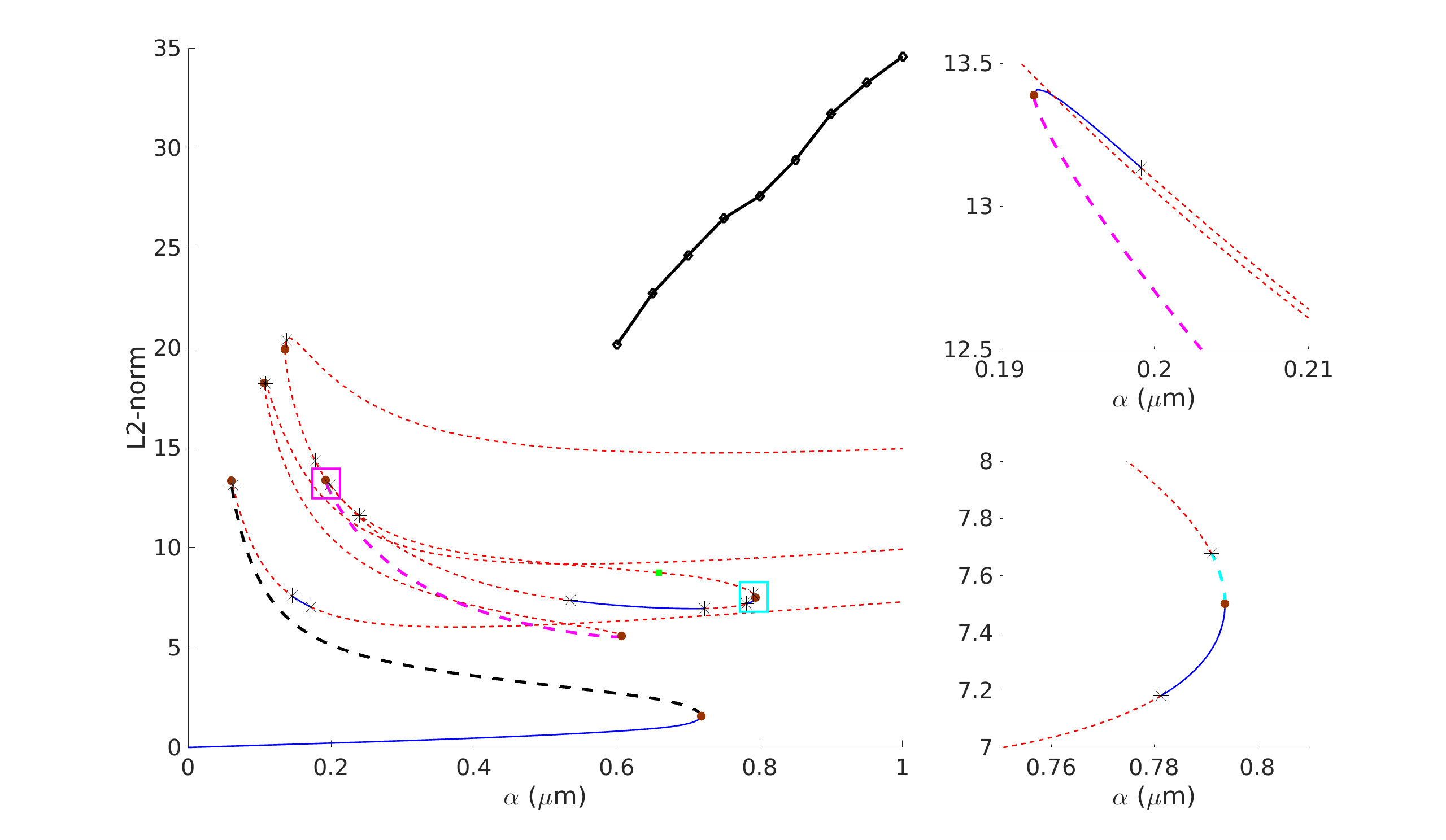}
\caption{A more complete bifurcation diagram and a zoom-in of the regions where stable branches appear. The color coding is same as before, and period-doubling bifurcations are shown as green squares. Two branches of saddle points with one-dimensional, unstable manifolds are colored
pink and cyan for easier reference.}
\label{fig:fullBif}
\end{figure*}

Seeing the interactions of the manifold of the black saddle with what appears to be another invariant object, we hypothesized that it may be another saddle which can be found close to where the manifold becomes distorted.  In order to converge to new solutions, we defined
the optimization problem $\min_u \lVert  u(0) - u(T) \rVert^2$ and solved it using local search methods.
Indeed, we were able to converge to a previously undiscovered branch of true periodic solutions and, using continuation techniques we now attained an extended bifurcation diagram shown in Fig.~\ref{fig:fullBif}.
As we can see the new solution branches are mostly unstable,
although it is interesting to point out an interval of bistability
for $\alpha in the (approximate) interval (0.5,0.7)$ 
%{\bf please refine}. 
Indeed, a stable
periodic orbit portion appears
to newly exist even in the vicinity of $\alpha \approx 0.79$, an interval
previously considered to be fully chaotic. In any event, the newly
identified portions of the bifurcation diagram
are not connected to the previous branches as can be seen
in Fig.~\ref{fig:fullBif}, at least in the parameter range we studied.
In total we find that there are three short, stable intervals; the longest one can be easily seen in Fig.~\ref{fig:fullBif}a surrounded by two NS bifurcations, similar
to the stable segment on the previously found branch. The two other stable branches are marked by the pink and cyan square, and the zoomed in bifurcation diagram is shown in Fig.~\ref{fig:fullBif}b,c respectively. Of particular interest are the unstable branches shown in pink and cyan color,
since the fixed points there are saddles with only a single unstable eigenvalue, hence the unstable manifold there can be easily computed. All other unstable branches have at least two unstable eigenvalues.
The pink and the cyan manifolds will be the object of study in the remainder of this work and will illustrate additional global bifurcations and how some of them may lead to the onset of chaos in this system.

%{\it More torus bifurcations.}

\subsection{More torus bifurcations}

\begin{figure*}[htbp]
\centering
\includegraphics[width=\linewidth]{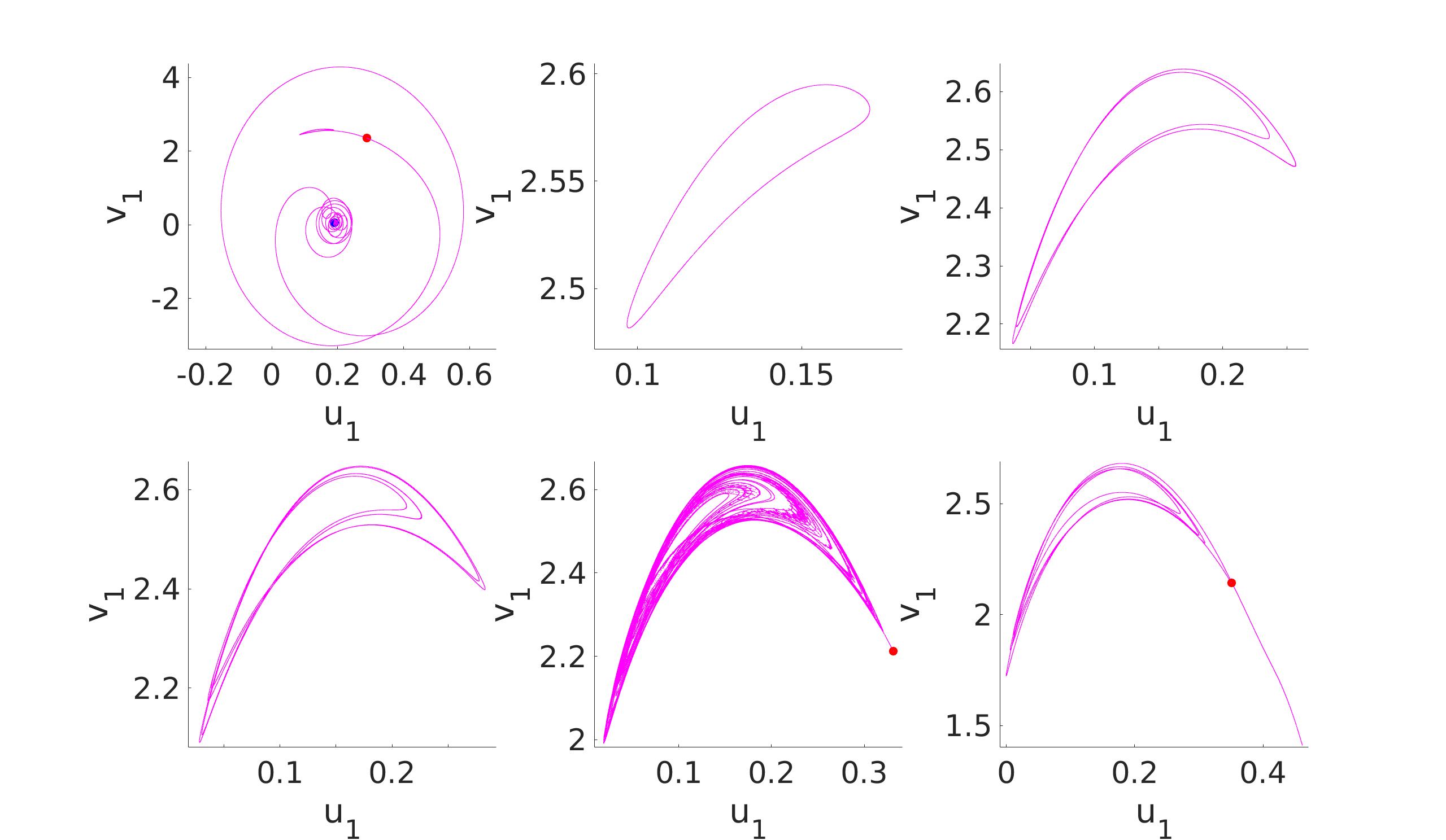}
\caption{Sequence of period-doubling bifurcations of the second torus and its eventual destruction. (a) $\alpha = 0.2$ $\mu$m, one side of the unstable manifold is attracted to the invariant circle and the other side to the low, stable branch (blue dot).
(b) $\alpha = 0.2$ $\mu$m, the invariant circle. (c) $\alpha = 0.2065$ $\mu$m, first period-doubling bifurcation. (d) $\alpha = 0.208$ $\mu$m, second period-doubling bifurcation. (e) $\alpha = 0.2095$ $\mu$m, no discernible invariant curve exists anymore.
(f) $\alpha = 0.215$ $\mu$m, the side of the manifold previously attracted to the torus is eventually attracted to the stable branch as well. All transitions are completely analogous to the transitions of the previous torus in Fig.~\ref{fig:firstTorus}.}
\label{fig:secTorus}
\end{figure*}

We begin the exploration of the new unstable manifolds at the top peak of the pink branch shown in Fig.~\ref{fig:fullBif}b. Initially, while the short stable branch still exists, one side of the unstable pink manifold is attracted to it. The other side
of the pink manifold is always quickly attracted to the lowest stable branch from the original bifurcation diagram unless specified otherwise. After the NS bifurcation a stable torus appears and is one of the two attractors (Fig.~\ref{fig:secTorus}a).
The final invariant circle can be seen in Fig.~\ref{fig:secTorus}b. Similar to the transitions of the black torus, the pink torus also undergoes a period-doubling bifurcation where it loses its smoothness and the invariant curve starts separating until it has a distinct double loop (Fig.~\ref{fig:secTorus}c).
A second period-doubling bifurcation occurs and the curve now has a quadruple loop (Fig.~\ref{fig:secTorus}d). While it was hard to determine additional period-doublings of the torus, eventually the torus is destroyed and no clear attractor exists anymore. The manifold wanders around the same region
of the phase space, where the torus existed (Fig.~\ref{fig:secTorus}e) in a chaotic transient. Increasing the amplitude further, the side of the manifold attracted to the torus becomes eventually attracted to the lowest stable branch as well (Fig.~\ref{fig:secTorus}f).
It is evident that the global bifurcations in the neighborhood of the pink torus and the local structure of solution branches and bifurcations is almost identical to the ones in the neighborhood of the black torus from the original bifurcation diagram.

\begin{figure*}[htbp]
\centering
\includegraphics[width=\linewidth]{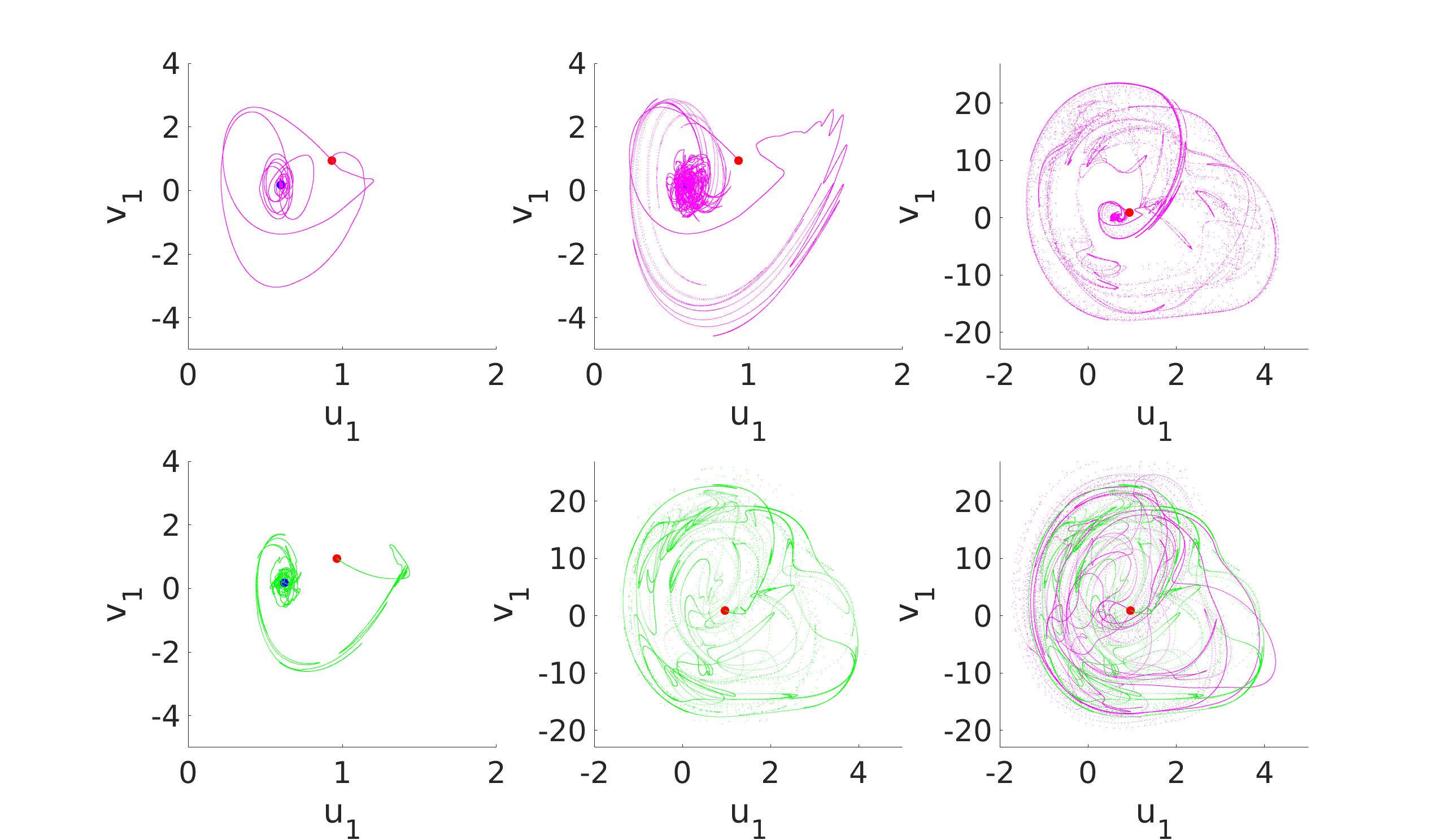}
\caption{Tangles of the pink manifold. (a) $\alpha = 0.571$ $\mu$m, the side of the manifold previously attracted to the torus. (b) $\alpha = 0.5734$ $\mu$m, the manifold becomes distorted due to heteroclinic tangencies, potentially with the stable manifold of the black saddle.
(c) $\alpha = 0.5735$ $\mu$m, the volume of the phase space visited by the pink manifold has expanded. It is now in the same region as the entangled black manifold. (d) $\alpha = 0.5847$ $\mu$m, the second side of the unstable manifold of the pink saddle exhibits heteroclinic tangencies as well.
This is shown in green for clarity. (e) $\alpha = 0.5848$ $\mu$m, the second side of the pink manifold is now entangled as well. (f) $\alpha = 0.5848$ $\mu$m, both sides of the pink manifold are shown.}
\label{fig:pinkSaddle1}
\end{figure*}

%{\it Pink manifold tangles.}
After the torus has been destroyed, both sides of the unstable pink manifold become ``smooth'' and are quickly attracted to the lowest stable branch. Fig.~\ref{fig:pinkSaddle1}a, for clarity, shows only the side of the manifold originally attracted to the torus.
This picture changes as we move to the right of the bifurcation diagram, where at some parameter value the same side of the manifold starts becoming distorted until it starts being reminiscent of manifold tangencies as well. If we compare the spatial profile of the granular chain at the point of these tangencies to
the spatial profile of the black saddle at the same amplitude value, we see that they match quite closely. This implies that the tangencies of the pink manifold are heteroclinic, where the unstable pink manifold becomes tangent to the stable manifold of the black saddle.
A slight increase in amplitude leads to manifold tangling and causes the phase space volume visited by the manifold to explode. The pink manifold is now attracted to the same transient chaos that was born from the homoclinic tangle of the black saddle.
This can be seen as additional proof that the tangencies are indeed heteroclinic. However, at very long times the only attractor is still the lowest stable branch.

After this the two manifolds become untangled leading to a picture similar to the one in Fig.~\ref{fig:pinkSaddle1}a until another tangency happens around the amplitude value $\alpha \approx 0.573$ $\mu$m.
This second transition is shown in Fig.~\ref{fig:pinkSaddle1}. Fig.~\ref{fig:pinkSaddle1}b shows the heteroclinic tangencies and Fig.~\ref{fig:pinkSaddle1}c shows the transient chaos visited by the pink manifold after the tangle.
Past this point the side of the pink manifold that was originally attracted to the torus remains entangled with the black manifold.

So far we've looked only at one side of the pink unstable manifold as the other was always attracted to the lowest stable branch. This is no longer true around the amplitude $\alpha \approx 0.584$ $\mu$m.
%{\bf question to be erased: can we explain this at all further, given that
%  the bottom stable branch is nowhere near disappearance at this value of
%  $\alpha$? It seems like a fair (potential referee) question...I tried
%  to write sth about it}
We connect this threshold  below to the emergence (and attractivity) of the
chaotic state past this value of $\alpha$ associated with the .
There, the other side of the pink manifold becomes tangent
to the black manifold as well (Fig.~\ref{fig:pinkSaddle1}d). It is shown in green here, to discern the two sides easier. Fig.~\ref{fig:pinkSaddle1}e shows the same side of the pink manifold {\em after the tangle with the black manifold}. Again, we can see that it is now
attracted to the transient chaos we observed previously. Fig.~\ref{fig:pinkSaddle1}f shows both sides of the pink manifold (one side in pink and the other in green) entangled with the black manifold (not shown) and being attracted to the large  chaotic region.
After this point the manifold remains entangled all the way until the pink saddle disappears in a saddle-node bifurcation around the amplitude $\alpha_{sn}^3 \approx 0.61$ \si{\mu {\rm m}}.
% Future consideration: spatial profile 
% of these apparently bulk states (emerging
% presumably from \alpha -> 0 as the dissipation
% vanishes. 

It is important to note that the entanglement of both sides of the pink manifold happens at an amplitude value that coincides with the value where the chaotic attractor appears to become stable - meaning that a state on the chaotic attractor remains there indefinitely
without being attracted to the lowest stable branch. 
We postulate that this transition makes the basin of attraction of the bottom entrained periodic stable branch very small, so that almost all states in the phase space are attracted to the chaotic region without ever crossing the chaotic attractor's basin boundary. 
The appearance/disappearance of large attractors has been traditionally studied under the term {\em crises/intermittency}, involving (as in our case) global interactions of invariant manifolds (e.g. \cite{A1,A2} and for a more recent example \cite{A3}).

\begin{figure}[htbp]
\centering
\includegraphics[width=\linewidth]{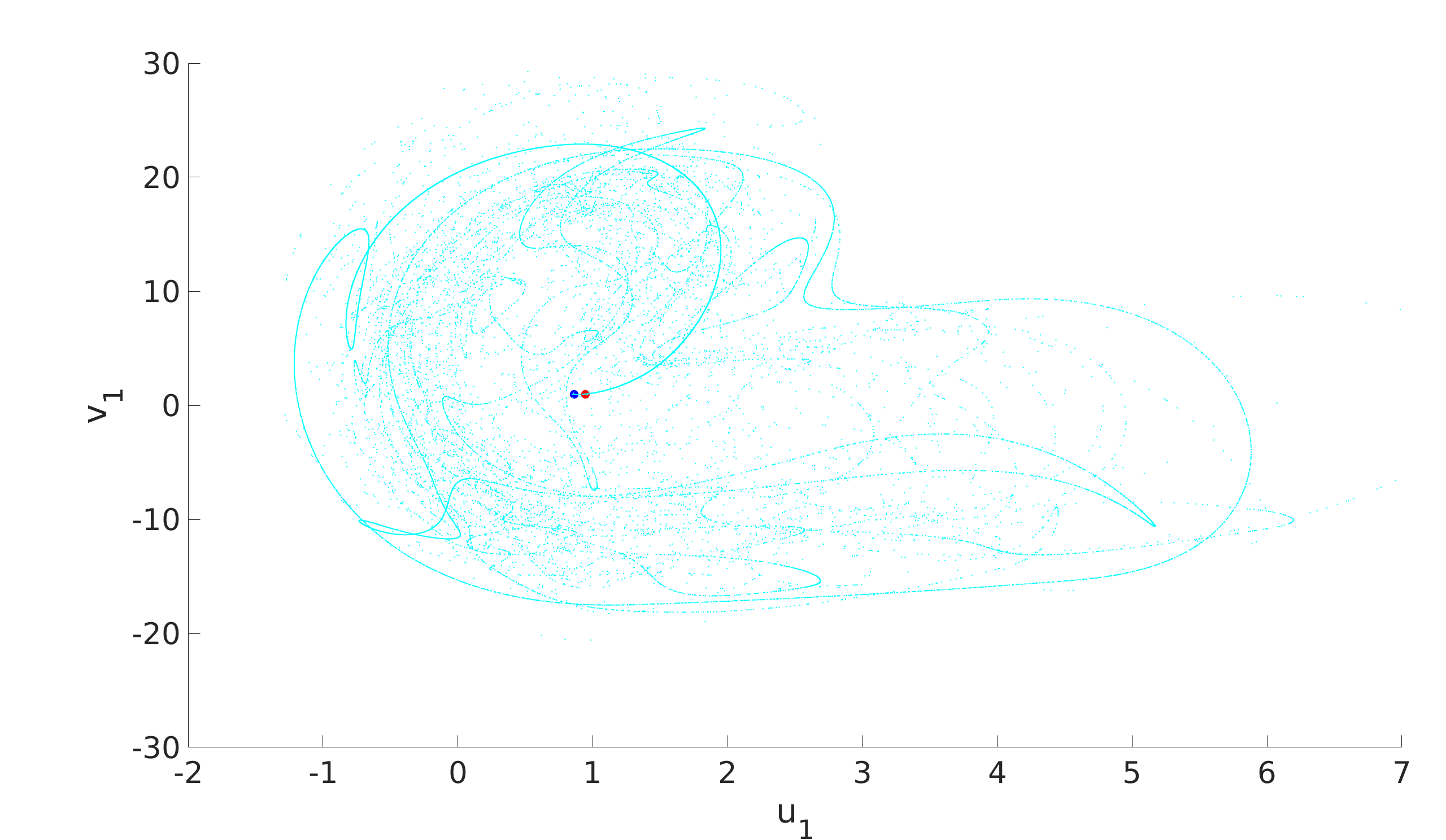}
\caption{Cyan manifold. $\alpha = 0.793$ $\mu$m, one side of the manifold is attracted to the same attractor as the black and pink manifolds, while
  the other to the stable blue branch that exists for this parameter value.}
\label{fig:cyanSaddle1}
\end{figure}

%{\it Cyan manifold.}
The last unstable branch we can study is colored in cyan in Fig.~\ref{fig:fullBif}c. This is again a branch of saddles with only a single unstable eigenvalue, whose unstable manifold we can compute. A sample of both sides of this manifold is shown
in Fig.~\ref{fig:cyanSaddle1}. One side of the manifold is attracted to the short stable branch that was born in the saddle-node bifurcation and is shown as a blue dot in the figure.
It is interesting to reiterate that this stable portion and its attractivity
had been missed earlier, e.g., in~\cite{hoogeboom2013hysteresis}.
The other side of the unstable manifold is attracted to the same chaotic attractor that we saw before, 
which still exists, even though the saddle, whose homoclinic tangle created it, has long disappeared (!) through the earlier saddle-node bifurcation
at $\alpha_{sn}^1$.

%{\bf Conclusions.}
\section{Conclusions and Future Challenges}

The nonlinear dynamics and bifurcations of a one-dimensional,
statically compressed, experimentally accessible diatomic granular crystal
with external forcing were examined in the present work. Previous works had found stable periodic solutions, along with chaotic responses and hysteretic dynamics.
In this work we expanded on the previous results using 
the computation of unstable manifolds of saddle-type periodic solutions (saddle stroboscopic points) as our main numerical tool. Tracking of these manifolds can reveal global bifurcations in the system and explain some of the dynamical phenomena observed.
For a certain range of forcing amplitudes the system possesses multiple
stable attractors, one periodic and one quasiperiodic; notice that
we also revealed novel scenarios with multiple periodic attractors.
The first global
bifurcation observed in the system is the destruction of a $T^2$ torus
through a cascade of period-doubling bifurcations.
Afterwards, only a periodic attractor exists, corresponding to a state of low energy propagation through the system. The second global bifurcation is observed at higher amplitude values and involves homoclinic tangles between the stable and unstable manifold
of a saddle point. After the manifolds have become entangled, the unstable manifold seems to pass through a region of transient chaos before re-converging to the aforementioned stable, periodic solution.

New branches  of solutions were uncovered, seeded with initial guesses from optimization techniques, motivated by dynamical observations. 
The newly discovered states are not continuously connected to the previously known ones
and possess both stable and unstable segments. Among the unstable segments some correspond to (stroboscopic) saddle points
with one-dimensional, unstable manifolds. The computation of these manifolds unveiled another destruction of a $T^2$ torus through a period-doubling cascade, completely analogous to the sequence of the first torus. In addition, the unstable
manifold of one of the new saddles exhibits heteroclinic tangencies with the manifolds of the previous saddle. We postulate that these entanglements resulting in a stable, full-blown chaotic attractor with a large basin. These results provide insight into the appearance of the chaotic response and show how the examination of global manifolds can suggest the existence of previously unknown branches of solutions and interactions between them. This confirms that the computation
of invariant manifolds in a dynamical system such as granular crystals is an important tool if one wants to obtain a more complete picture of the dynamics, and should be taken into consideration when trying to design a granular system with specified properties.

Naturally, it would be relevant to expand upon the usage of such
tools in damped-driven many degree-of-freedom dynamical systems.
Recently experiments (and theoretical/numerical analysis) were
extended to spatially two-dimensional systems of the same general
class (albeit with repelling magnets rather than granular chains)
in two-dimensional configurations including, e.g., light-mass
defects. It may be interesting to envision such two-dimensional
analogues of damped-driven engineered granular crystals as canonical
testbeds where similar ideas can be extended to understand the
role of dimensionality as concerns the potential system states
and the wealth of dynamical phenomena within such higher-dimensional
case examples. Such studies are currently under consideration and will
be reported in future publications.

\newpage

\bibliographystyle{apsrev4-1}
\bibliography{thesis_v2.bib}

\begin{thebibliography}{49}%
\makeatletter
\providecommand \@ifxundefined [1]{%
 \@ifx{#1\undefined}
}%
\providecommand \@ifnum [1]{%
 \ifnum #1\expandafter \@firstoftwo
 \else \expandafter \@secondoftwo
 \fi
}%
\providecommand \@ifx [1]{%
 \ifx #1\expandafter \@firstoftwo
 \else \expandafter \@secondoftwo
 \fi
}%
\providecommand \natexlab [1]{#1}%
\providecommand \enquote  [1]{``#1''}%
\providecommand \bibnamefont  [1]{#1}%
\providecommand \bibfnamefont [1]{#1}%
\providecommand \citenamefont [1]{#1}%
\providecommand \href@noop [0]{\@secondoftwo}%
\providecommand \href [0]{\begingroup \@sanitize@url \@href}%
\providecommand \@href[1]{\@@startlink{#1}\@@href}%
\providecommand \@@href[1]{\endgroup#1\@@endlink}%
\providecommand \@sanitize@url [0]{\catcode `\\12\catcode `\$12\catcode
  `\&12\catcode `\#12\catcode `\^12\catcode `\_12\catcode `\%12\relax}%
\providecommand \@@startlink[1]{}%
\providecommand \@@endlink[0]{}%
\providecommand \url  [0]{\begingroup\@sanitize@url \@url }%
\providecommand \@url [1]{\endgroup\@href {#1}{\urlprefix }}%
\providecommand \urlprefix  [0]{URL }%
\providecommand \Eprint [0]{\href }%
\providecommand \doibase [0]{http://dx.doi.org/}%
\providecommand \selectlanguage [0]{\@gobble}%
\providecommand \bibinfo  [0]{\@secondoftwo}%
\providecommand \bibfield  [0]{\@secondoftwo}%
\providecommand \translation [1]{[#1]}%
\providecommand \BibitemOpen [0]{}%
\providecommand \bibitemStop [0]{}%
\providecommand \bibitemNoStop [0]{.\EOS\space}%
\providecommand \EOS [0]{\spacefactor3000\relax}%
\providecommand \BibitemShut  [1]{\csname bibitem#1\endcsname}%
\let\auto@bib@innerbib\@empty
%</preamble>
\bibitem [{\citenamefont {Nesterenko}(2001)}]{nesterenko2001dynamics}%
  \BibitemOpen
  \bibfield  {author} {\bibinfo {author} {\bibfnamefont {V.}~\bibnamefont
  {Nesterenko}},\ }\href@noop {} {\emph {\bibinfo {title} {Dynamics of
  heterogeneous materials}}}\ (\bibinfo  {publisher} {Springer-Verlag},\
  \bibinfo {year} {2001})\BibitemShut {NoStop}%
\bibitem [{\citenamefont {Sen}\ \emph {et~al.}(2008)\citenamefont {Sen},
  \citenamefont {Hong}, \citenamefont {Bang}, \citenamefont {Avalos},\ and\
  \citenamefont {Doney}}]{sen2008solitary}%
  \BibitemOpen
  \bibfield  {author} {\bibinfo {author} {\bibfnamefont {S.}~\bibnamefont
  {Sen}}, \bibinfo {author} {\bibfnamefont {J.}~\bibnamefont {Hong}}, \bibinfo
  {author} {\bibfnamefont {J.}~\bibnamefont {Bang}}, \bibinfo {author}
  {\bibfnamefont {E.}~\bibnamefont {Avalos}}, \ and\ \bibinfo {author}
  {\bibfnamefont {R.}~\bibnamefont {Doney}},\ }\href@noop {} {\bibfield
  {journal} {\bibinfo  {journal} {Physics Reports}\ }\textbf {\bibinfo {volume}
  {462}},\ \bibinfo {pages} {21} (\bibinfo {year} {2008})}\BibitemShut
  {NoStop}%
\bibitem [{\citenamefont {Vakakis}(2012)}]{vakakis_review}%
  \BibitemOpen
  \bibfield  {author} {\bibinfo {author} {\bibfnamefont {A.~F.}\ \bibnamefont
  {Vakakis}},\ }in\ \href@noop {} {\emph {\bibinfo {booktitle} {Wave
  Propagation in Linear and Nonlinear Periodic Media (International Center for
  Mechanical Sciences (CISM) Courses and Lectures)}}}\ (\bibinfo  {publisher}
  {Springer-Verlag, Berlin, Germany},\ \bibinfo {year} {2012})\ p.\ \bibinfo
  {pages} {257}\BibitemShut {NoStop}%
\bibitem [{\citenamefont {Starosvetsky}\ \emph {et~al.}(2017)\citenamefont
  {Starosvetsky}, \citenamefont {Jayaprakash}, \citenamefont {Hasan},\ and\
  \citenamefont {Vakakis}}]{yuli_book}%
  \BibitemOpen
  \bibfield  {author} {\bibinfo {author} {\bibfnamefont {Y.}~\bibnamefont
  {Starosvetsky}}, \bibinfo {author} {\bibfnamefont {K.}~\bibnamefont
  {Jayaprakash}}, \bibinfo {author} {\bibfnamefont {M.~A.}\ \bibnamefont
  {Hasan}}, \ and\ \bibinfo {author} {\bibfnamefont {A.}~\bibnamefont
  {Vakakis}},\ }\href@noop {} {\emph {\bibinfo {title} {Dynamics and Acoustics
  of Ordered Granular Media}}}\ (\bibinfo  {publisher} {World Scientific,
  Singapore},\ \bibinfo {year} {2017})\BibitemShut {NoStop}%
\bibitem [{\citenamefont {Chong}\ and\ \citenamefont
  {Kevrekidis}(2018)}]{granularBook}%
  \BibitemOpen
  \bibfield  {author} {\bibinfo {author} {\bibfnamefont {C.}~\bibnamefont
  {Chong}}\ and\ \bibinfo {author} {\bibfnamefont {P.~G.}\ \bibnamefont
  {Kevrekidis}},\ }\href@noop {} {\emph {\bibinfo {title} {Coherent Structures
  in Granular Crystals: From Experiment and Modelling to Computation and
  Mathematical Analysis}}}\ (\bibinfo  {publisher} {Springer},\ \bibinfo
  {address} {New York},\ \bibinfo {year} {2018})\BibitemShut {NoStop}%
\bibitem [{\citenamefont {Nesterenko}(1983)}]{nesterenko1983propagation}%
  \BibitemOpen
  \bibfield  {author} {\bibinfo {author} {\bibfnamefont {V.}~\bibnamefont
  {Nesterenko}},\ }\href@noop {} {\bibfield  {journal} {\bibinfo  {journal}
  {Journal of Applied Mechanics and Technical Physics}\ }\textbf {\bibinfo
  {volume} {24}},\ \bibinfo {pages} {733} (\bibinfo {year} {1983})}\BibitemShut
  {NoStop}%
\bibitem [{\citenamefont {Lazaridi}\ and\ \citenamefont
  {Nesterenko}(1985)}]{lazaridi1985observation}%
  \BibitemOpen
  \bibfield  {author} {\bibinfo {author} {\bibfnamefont {A.}~\bibnamefont
  {Lazaridi}}\ and\ \bibinfo {author} {\bibfnamefont {V.}~\bibnamefont
  {Nesterenko}},\ }\href@noop {} {\bibfield  {journal} {\bibinfo  {journal}
  {Journal of Applied Mechanics and Technical Physics}\ }\textbf {\bibinfo
  {volume} {26}},\ \bibinfo {pages} {405} (\bibinfo {year} {1985})}\BibitemShut
  {NoStop}%
\bibitem [{\citenamefont {Coste}\ \emph {et~al.}(1997)\citenamefont {Coste},
  \citenamefont {Falcon},\ and\ \citenamefont {Fauve}}]{coste1997solitary}%
  \BibitemOpen
  \bibfield  {author} {\bibinfo {author} {\bibfnamefont {C.}~\bibnamefont
  {Coste}}, \bibinfo {author} {\bibfnamefont {E.}~\bibnamefont {Falcon}}, \
  and\ \bibinfo {author} {\bibfnamefont {S.}~\bibnamefont {Fauve}},\
  }\href@noop {} {\bibfield  {journal} {\bibinfo  {journal} {Physical review
  E}\ }\textbf {\bibinfo {volume} {56}},\ \bibinfo {pages} {6104} (\bibinfo
  {year} {1997})}\BibitemShut {NoStop}%
\bibitem [{\citenamefont {Daraio}\ \emph
  {et~al.}(2006{\natexlab{a}})\citenamefont {Daraio}, \citenamefont
  {Nesterenko}, \citenamefont {Herbold},\ and\ \citenamefont
  {Jin}}]{daraio2006tunability}%
  \BibitemOpen
  \bibfield  {author} {\bibinfo {author} {\bibfnamefont {C.}~\bibnamefont
  {Daraio}}, \bibinfo {author} {\bibfnamefont {V.}~\bibnamefont {Nesterenko}},
  \bibinfo {author} {\bibfnamefont {E.}~\bibnamefont {Herbold}}, \ and\
  \bibinfo {author} {\bibfnamefont {S.}~\bibnamefont {Jin}},\ }\href@noop {}
  {\bibfield  {journal} {\bibinfo  {journal} {Physical Review E}\ }\textbf
  {\bibinfo {volume} {73}},\ \bibinfo {pages} {026610} (\bibinfo {year}
  {2006}{\natexlab{a}})}\BibitemShut {NoStop}%
\bibitem [{\citenamefont {Porter}\ \emph {et~al.}(2008)\citenamefont {Porter},
  \citenamefont {Daraio}, \citenamefont {Herbold}, \citenamefont
  {Szelengowicz},\ and\ \citenamefont {Kevrekidis}}]{porter2008highly}%
  \BibitemOpen
  \bibfield  {author} {\bibinfo {author} {\bibfnamefont {M.~A.}\ \bibnamefont
  {Porter}}, \bibinfo {author} {\bibfnamefont {C.}~\bibnamefont {Daraio}},
  \bibinfo {author} {\bibfnamefont {E.~B.}\ \bibnamefont {Herbold}}, \bibinfo
  {author} {\bibfnamefont {I.}~\bibnamefont {Szelengowicz}}, \ and\ \bibinfo
  {author} {\bibfnamefont {P.}~\bibnamefont {Kevrekidis}},\ }\href@noop {}
  {\bibfield  {journal} {\bibinfo  {journal} {Physical Review E}\ }\textbf
  {\bibinfo {volume} {77}},\ \bibinfo {pages} {015601} (\bibinfo {year}
  {2008})}\BibitemShut {NoStop}%
\bibitem [{\citenamefont {Starosvetsky}\ and\ \citenamefont
  {Vakakis}(2010)}]{starosvetsky2010traveling}%
  \BibitemOpen
  \bibfield  {author} {\bibinfo {author} {\bibfnamefont {Y.}~\bibnamefont
  {Starosvetsky}}\ and\ \bibinfo {author} {\bibfnamefont {A.~F.}\ \bibnamefont
  {Vakakis}},\ }\href@noop {} {\bibfield  {journal} {\bibinfo  {journal}
  {Physical Review E}\ }\textbf {\bibinfo {volume} {82}},\ \bibinfo {pages}
  {026603} (\bibinfo {year} {2010})}\BibitemShut {NoStop}%
\bibitem [{\citenamefont {Jayaprakash}\ \emph {et~al.}(2011)\citenamefont
  {Jayaprakash}, \citenamefont {Starosvetsky},\ and\ \citenamefont
  {Vakakis}}]{jayaprakash2011new}%
  \BibitemOpen
  \bibfield  {author} {\bibinfo {author} {\bibfnamefont {K.}~\bibnamefont
  {Jayaprakash}}, \bibinfo {author} {\bibfnamefont {Y.}~\bibnamefont
  {Starosvetsky}}, \ and\ \bibinfo {author} {\bibfnamefont {A.~F.}\
  \bibnamefont {Vakakis}},\ }\href@noop {} {\bibfield  {journal} {\bibinfo
  {journal} {Physical Review E}\ }\textbf {\bibinfo {volume} {83}},\ \bibinfo
  {pages} {036606} (\bibinfo {year} {2011})}\BibitemShut {NoStop}%
\bibitem [{\citenamefont {Boechler}\ \emph {et~al.}(2010)\citenamefont
  {Boechler}, \citenamefont {Theocharis}, \citenamefont {Job}, \citenamefont
  {Kevrekidis}, \citenamefont {Porter},\ and\ \citenamefont
  {Daraio}}]{boechler2010discrete}%
  \BibitemOpen
  \bibfield  {author} {\bibinfo {author} {\bibfnamefont {N.}~\bibnamefont
  {Boechler}}, \bibinfo {author} {\bibfnamefont {G.}~\bibnamefont
  {Theocharis}}, \bibinfo {author} {\bibfnamefont {S.}~\bibnamefont {Job}},
  \bibinfo {author} {\bibfnamefont {P.}~\bibnamefont {Kevrekidis}}, \bibinfo
  {author} {\bibfnamefont {M.~A.}\ \bibnamefont {Porter}}, \ and\ \bibinfo
  {author} {\bibfnamefont {C.}~\bibnamefont {Daraio}},\ }\href@noop {}
  {\bibfield  {journal} {\bibinfo  {journal} {Physical review letters}\
  }\textbf {\bibinfo {volume} {104}},\ \bibinfo {pages} {244302} (\bibinfo
  {year} {2010})}\BibitemShut {NoStop}%
\bibitem [{\citenamefont {Theocharis}\ \emph {et~al.}(2010)\citenamefont
  {Theocharis}, \citenamefont {Boechler}, \citenamefont {Kevrekidis},
  \citenamefont {Job}, \citenamefont {Porter},\ and\ \citenamefont
  {Daraio}}]{theocharis2010intrinsic}%
  \BibitemOpen
  \bibfield  {author} {\bibinfo {author} {\bibfnamefont {G.}~\bibnamefont
  {Theocharis}}, \bibinfo {author} {\bibfnamefont {N.}~\bibnamefont
  {Boechler}}, \bibinfo {author} {\bibfnamefont {P.}~\bibnamefont
  {Kevrekidis}}, \bibinfo {author} {\bibfnamefont {S.}~\bibnamefont {Job}},
  \bibinfo {author} {\bibfnamefont {M.~A.}\ \bibnamefont {Porter}}, \ and\
  \bibinfo {author} {\bibfnamefont {C.}~\bibnamefont {Daraio}},\ }\href@noop {}
  {\bibfield  {journal} {\bibinfo  {journal} {Physical Review E}\ }\textbf
  {\bibinfo {volume} {82}},\ \bibinfo {pages} {056604} (\bibinfo {year}
  {2010})}\BibitemShut {NoStop}%
\bibitem [{\citenamefont {Chong}\ \emph {et~al.}(2014)\citenamefont {Chong},
  \citenamefont {Li}, \citenamefont {Yang}, \citenamefont {Williams},
  \citenamefont {Kevrekidis}, \citenamefont {Kevrekidis},\ and\ \citenamefont
  {Daraio}}]{chong2014damped}%
  \BibitemOpen
  \bibfield  {author} {\bibinfo {author} {\bibfnamefont {C.}~\bibnamefont
  {Chong}}, \bibinfo {author} {\bibfnamefont {F.}~\bibnamefont {Li}}, \bibinfo
  {author} {\bibfnamefont {J.}~\bibnamefont {Yang}}, \bibinfo {author}
  {\bibfnamefont {M.}~\bibnamefont {Williams}}, \bibinfo {author}
  {\bibfnamefont {I.}~\bibnamefont {Kevrekidis}}, \bibinfo {author}
  {\bibfnamefont {P.}~\bibnamefont {Kevrekidis}}, \ and\ \bibinfo {author}
  {\bibfnamefont {C.}~\bibnamefont {Daraio}},\ }\href@noop {} {\bibfield
  {journal} {\bibinfo  {journal} {Physical Review E}\ }\textbf {\bibinfo
  {volume} {89}},\ \bibinfo {pages} {032924} (\bibinfo {year}
  {2014})}\BibitemShut {NoStop}%
\bibitem [{\citenamefont {Daraio}\ \emph
  {et~al.}(2006{\natexlab{b}})\citenamefont {Daraio}, \citenamefont
  {Nesterenko}, \citenamefont {Herbold},\ and\ \citenamefont
  {Jin}}]{daraio2006energy}%
  \BibitemOpen
  \bibfield  {author} {\bibinfo {author} {\bibfnamefont {C.}~\bibnamefont
  {Daraio}}, \bibinfo {author} {\bibfnamefont {V.}~\bibnamefont {Nesterenko}},
  \bibinfo {author} {\bibfnamefont {E.}~\bibnamefont {Herbold}}, \ and\
  \bibinfo {author} {\bibfnamefont {S.}~\bibnamefont {Jin}},\ }\href@noop {}
  {\bibfield  {journal} {\bibinfo  {journal} {Physical Review Letters}\
  }\textbf {\bibinfo {volume} {96}},\ \bibinfo {pages} {058002} (\bibinfo
  {year} {2006}{\natexlab{b}})}\BibitemShut {NoStop}%
\bibitem [{\citenamefont {Herbold}\ and\ \citenamefont
  {Nesterenko}(2007)}]{herbold2007shock}%
  \BibitemOpen
  \bibfield  {author} {\bibinfo {author} {\bibfnamefont {E.}~\bibnamefont
  {Herbold}}\ and\ \bibinfo {author} {\bibfnamefont {V.}~\bibnamefont
  {Nesterenko}},\ }\href@noop {} {\bibfield  {journal} {\bibinfo  {journal}
  {Physical Review E}\ }\textbf {\bibinfo {volume} {75}},\ \bibinfo {pages}
  {021304} (\bibinfo {year} {2007})}\BibitemShut {NoStop}%
\bibitem [{\citenamefont {Molinari}\ and\ \citenamefont
  {Daraio}(2009)}]{molinari2009stationary}%
  \BibitemOpen
  \bibfield  {author} {\bibinfo {author} {\bibfnamefont {A.}~\bibnamefont
  {Molinari}}\ and\ \bibinfo {author} {\bibfnamefont {C.}~\bibnamefont
  {Daraio}},\ }\href@noop {} {\bibfield  {journal} {\bibinfo  {journal}
  {Physical Review E}\ }\textbf {\bibinfo {volume} {80}},\ \bibinfo {pages}
  {056602} (\bibinfo {year} {2009})}\BibitemShut {NoStop}%
\bibitem [{\citenamefont {Chong}\ \emph {et~al.}(2024)\citenamefont {Chong},
  \citenamefont {Geisler}, \citenamefont {Kevrekidis},\ and\ \citenamefont
  {Biondini}}]{KdVToda_limits}%
  \BibitemOpen
  \bibfield  {author} {\bibinfo {author} {\bibfnamefont {C.}~\bibnamefont
  {Chong}}, \bibinfo {author} {\bibfnamefont {A.}~\bibnamefont {Geisler}},
  \bibinfo {author} {\bibfnamefont {P.}~\bibnamefont {Kevrekidis}}, \ and\
  \bibinfo {author} {\bibfnamefont {G.}~\bibnamefont {Biondini}},\ }\href@noop
  {} {\bibfield  {journal} {\bibinfo  {journal} {arXiv:2402.08218}\ } (\bibinfo
  {year} {2024})}\BibitemShut {NoStop}%
\bibitem [{\citenamefont {Li}\ \emph {et~al.}(2021)\citenamefont {Li},
  \citenamefont {Chockalingam},\ and\ \citenamefont {Cohen}}]{talcohen}%
  \BibitemOpen
  \bibfield  {author} {\bibinfo {author} {\bibfnamefont {J.}~\bibnamefont
  {Li}}, \bibinfo {author} {\bibfnamefont {S.}~\bibnamefont {Chockalingam}}, \
  and\ \bibinfo {author} {\bibfnamefont {T.}~\bibnamefont {Cohen}},\ }\href
  {\doibase 10.1103/PhysRevLett.127.014302} {\bibfield  {journal} {\bibinfo
  {journal} {Phys. Rev. Lett.}\ }\textbf {\bibinfo {volume} {127}},\ \bibinfo
  {pages} {014302} (\bibinfo {year} {2021})}\BibitemShut {NoStop}%
\bibitem [{\citenamefont {Pozharskiy}\ \emph {et~al.}(2015)\citenamefont
  {Pozharskiy}, \citenamefont {Zhang}, \citenamefont {Williams}, \citenamefont
  {McFarland}, \citenamefont {Kevrekidis}, \citenamefont {Vakakis},\ and\
  \citenamefont {Kevrekidis}}]{pozharskiy2015nonlinear}%
  \BibitemOpen
  \bibfield  {author} {\bibinfo {author} {\bibfnamefont {D.}~\bibnamefont
  {Pozharskiy}}, \bibinfo {author} {\bibfnamefont {Y.}~\bibnamefont {Zhang}},
  \bibinfo {author} {\bibfnamefont {M.}~\bibnamefont {Williams}}, \bibinfo
  {author} {\bibfnamefont {D.}~\bibnamefont {McFarland}}, \bibinfo {author}
  {\bibfnamefont {P.~G.}\ \bibnamefont {Kevrekidis}}, \bibinfo {author}
  {\bibfnamefont {A.}~\bibnamefont {Vakakis}}, \ and\ \bibinfo {author}
  {\bibfnamefont {I.}~\bibnamefont {Kevrekidis}},\ }\href@noop {} {\bibfield
  {journal} {\bibinfo  {journal} {Physical Review E}\ }\textbf {\bibinfo
  {volume} {92}},\ \bibinfo {pages} {063203} (\bibinfo {year}
  {2015})}\BibitemShut {NoStop}%
\bibitem [{\citenamefont {Zhang}\ \emph {et~al.}(2017)\citenamefont {Zhang},
  \citenamefont {Pozharskiy}, \citenamefont {McFarland}, \citenamefont
  {Kevrekidis}, \citenamefont {Kevrekidis},\ and\ \citenamefont
  {Vakakis}}]{zhang2017experimental}%
  \BibitemOpen
  \bibfield  {author} {\bibinfo {author} {\bibfnamefont {Y.}~\bibnamefont
  {Zhang}}, \bibinfo {author} {\bibfnamefont {D.}~\bibnamefont {Pozharskiy}},
  \bibinfo {author} {\bibfnamefont {D.~M.}\ \bibnamefont {McFarland}}, \bibinfo
  {author} {\bibfnamefont {P.~G.}\ \bibnamefont {Kevrekidis}}, \bibinfo
  {author} {\bibfnamefont {I.~G.}\ \bibnamefont {Kevrekidis}}, \ and\ \bibinfo
  {author} {\bibfnamefont {A.~F.}\ \bibnamefont {Vakakis}},\ }\href@noop {}
  {\bibfield  {journal} {\bibinfo  {journal} {Experimental Mechanics}\ }\textbf
  {\bibinfo {volume} {57}},\ \bibinfo {pages} {505} (\bibinfo {year}
  {2017})}\BibitemShut {NoStop}%
\bibitem [{\citenamefont {Fraternali}\ \emph {et~al.}(2009)\citenamefont
  {Fraternali}, \citenamefont {Porter},\ and\ \citenamefont
  {Daraio}}]{fraternali2009optimal}%
  \BibitemOpen
  \bibfield  {author} {\bibinfo {author} {\bibfnamefont {F.}~\bibnamefont
  {Fraternali}}, \bibinfo {author} {\bibfnamefont {M.~A.}\ \bibnamefont
  {Porter}}, \ and\ \bibinfo {author} {\bibfnamefont {C.}~\bibnamefont
  {Daraio}},\ }\href@noop {} {\bibfield  {journal} {\bibinfo  {journal}
  {Mechanics of Advanced Materials and Structures}\ }\textbf {\bibinfo {volume}
  {17}},\ \bibinfo {pages} {1} (\bibinfo {year} {2009})}\BibitemShut {NoStop}%
\bibitem [{\citenamefont {Spadoni}\ and\ \citenamefont
  {Daraio}(2010)}]{spadoni2010generation}%
  \BibitemOpen
  \bibfield  {author} {\bibinfo {author} {\bibfnamefont {A.}~\bibnamefont
  {Spadoni}}\ and\ \bibinfo {author} {\bibfnamefont {C.}~\bibnamefont
  {Daraio}},\ }\href@noop {} {\bibfield  {journal} {\bibinfo  {journal}
  {Proceedings of the National Academy of Sciences}\ }\textbf {\bibinfo
  {volume} {107}},\ \bibinfo {pages} {7230} (\bibinfo {year}
  {2010})}\BibitemShut {NoStop}%
\bibitem [{\citenamefont {Khatri}\ \emph {et~al.}(2008)\citenamefont {Khatri},
  \citenamefont {Daraio},\ and\ \citenamefont {Rizzo}}]{khatri2008highly}%
  \BibitemOpen
  \bibfield  {author} {\bibinfo {author} {\bibfnamefont {D.}~\bibnamefont
  {Khatri}}, \bibinfo {author} {\bibfnamefont {C.}~\bibnamefont {Daraio}}, \
  and\ \bibinfo {author} {\bibfnamefont {P.}~\bibnamefont {Rizzo}}\ }(\bibinfo
  {organization} {Society of Photo-optical Instrumentation Engineers},\
  \bibinfo {year} {2008})\BibitemShut {NoStop}%
\bibitem [{\citenamefont {Nesterenko}\ \emph {et~al.}(2005)\citenamefont
  {Nesterenko}, \citenamefont {Daraio}, \citenamefont {Herbold},\ and\
  \citenamefont {Jin}}]{nesterenko2005anomalous}%
  \BibitemOpen
  \bibfield  {author} {\bibinfo {author} {\bibfnamefont {V.}~\bibnamefont
  {Nesterenko}}, \bibinfo {author} {\bibfnamefont {C.}~\bibnamefont {Daraio}},
  \bibinfo {author} {\bibfnamefont {E.}~\bibnamefont {Herbold}}, \ and\
  \bibinfo {author} {\bibfnamefont {S.}~\bibnamefont {Jin}},\ }\href@noop {}
  {\bibfield  {journal} {\bibinfo  {journal} {Physical review letters}\
  }\textbf {\bibinfo {volume} {95}},\ \bibinfo {pages} {158702} (\bibinfo
  {year} {2005})}\BibitemShut {NoStop}%
\bibitem [{\citenamefont {Daraio}\ \emph {et~al.}(2005)\citenamefont {Daraio},
  \citenamefont {Nesterenko}, \citenamefont {Herbold},\ and\ \citenamefont
  {Jin}}]{daraio2005strongly}%
  \BibitemOpen
  \bibfield  {author} {\bibinfo {author} {\bibfnamefont {C.}~\bibnamefont
  {Daraio}}, \bibinfo {author} {\bibfnamefont {V.}~\bibnamefont {Nesterenko}},
  \bibinfo {author} {\bibfnamefont {E.}~\bibnamefont {Herbold}}, \ and\
  \bibinfo {author} {\bibfnamefont {S.}~\bibnamefont {Jin}},\ }\href@noop {}
  {\bibfield  {journal} {\bibinfo  {journal} {Physical Review E}\ }\textbf
  {\bibinfo {volume} {72}},\ \bibinfo {pages} {016603} (\bibinfo {year}
  {2005})}\BibitemShut {NoStop}%
\bibitem [{\citenamefont {Li}\ \emph {et~al.}(2014)\citenamefont {Li},
  \citenamefont {Anzel}, \citenamefont {Yang}, \citenamefont {Kevrekidis},\
  and\ \citenamefont {Daraio}}]{li2014granular}%
  \BibitemOpen
  \bibfield  {author} {\bibinfo {author} {\bibfnamefont {F.}~\bibnamefont
  {Li}}, \bibinfo {author} {\bibfnamefont {P.}~\bibnamefont {Anzel}}, \bibinfo
  {author} {\bibfnamefont {J.}~\bibnamefont {Yang}}, \bibinfo {author}
  {\bibfnamefont {P.~G.}\ \bibnamefont {Kevrekidis}}, \ and\ \bibinfo {author}
  {\bibfnamefont {C.}~\bibnamefont {Daraio}},\ }\href@noop {} {\bibfield
  {journal} {\bibinfo  {journal} {Nature communications}\ }\textbf {\bibinfo
  {volume} {5}},\ \bibinfo {pages} {5311} (\bibinfo {year} {2014})}\BibitemShut
  {NoStop}%
\bibitem [{\citenamefont {Boechler}\ \emph {et~al.}(2011)\citenamefont
  {Boechler}, \citenamefont {Theocharis},\ and\ \citenamefont
  {Daraio}}]{boechler2011bifurcation}%
  \BibitemOpen
  \bibfield  {author} {\bibinfo {author} {\bibfnamefont {N.}~\bibnamefont
  {Boechler}}, \bibinfo {author} {\bibfnamefont {G.}~\bibnamefont
  {Theocharis}}, \ and\ \bibinfo {author} {\bibfnamefont {C.}~\bibnamefont
  {Daraio}},\ }\href@noop {} {\bibfield  {journal} {\bibinfo  {journal} {Nature
  materials}\ }\textbf {\bibinfo {volume} {10}},\ \bibinfo {pages} {665}
  (\bibinfo {year} {2011})}\BibitemShut {NoStop}%
\bibitem [{\citenamefont {Rosas}\ \emph {et~al.}(2007)\citenamefont {Rosas},
  \citenamefont {Romero}, \citenamefont {Nesterenko},\ and\ \citenamefont
  {Lindenberg}}]{rosas2007observation}%
  \BibitemOpen
  \bibfield  {author} {\bibinfo {author} {\bibfnamefont {A.}~\bibnamefont
  {Rosas}}, \bibinfo {author} {\bibfnamefont {A.~H.}\ \bibnamefont {Romero}},
  \bibinfo {author} {\bibfnamefont {V.~F.}\ \bibnamefont {Nesterenko}}, \ and\
  \bibinfo {author} {\bibfnamefont {K.}~\bibnamefont {Lindenberg}},\
  }\href@noop {} {\bibfield  {journal} {\bibinfo  {journal} {Physical review
  letters}\ }\textbf {\bibinfo {volume} {98}},\ \bibinfo {pages} {164301}
  (\bibinfo {year} {2007})}\BibitemShut {NoStop}%
\bibitem [{\citenamefont {Rosas}\ \emph {et~al.}(2008)\citenamefont {Rosas},
  \citenamefont {Romero}, \citenamefont {Nesterenko},\ and\ \citenamefont
  {Lindenberg}}]{rosas2008short}%
  \BibitemOpen
  \bibfield  {author} {\bibinfo {author} {\bibfnamefont {A.}~\bibnamefont
  {Rosas}}, \bibinfo {author} {\bibfnamefont {A.~H.}\ \bibnamefont {Romero}},
  \bibinfo {author} {\bibfnamefont {V.~F.}\ \bibnamefont {Nesterenko}}, \ and\
  \bibinfo {author} {\bibfnamefont {K.}~\bibnamefont {Lindenberg}},\
  }\href@noop {} {\bibfield  {journal} {\bibinfo  {journal} {Physical Review
  E}\ }\textbf {\bibinfo {volume} {78}},\ \bibinfo {pages} {051303} (\bibinfo
  {year} {2008})}\BibitemShut {NoStop}%
\bibitem [{\citenamefont {Carretero-Gonz{\'a}lez}\ \emph
  {et~al.}(2009)\citenamefont {Carretero-Gonz{\'a}lez}, \citenamefont {Khatri},
  \citenamefont {Porter}, \citenamefont {Kevrekidis},\ and\ \citenamefont
  {Daraio}}]{carretero2009dissipative}%
  \BibitemOpen
  \bibfield  {author} {\bibinfo {author} {\bibfnamefont {R.}~\bibnamefont
  {Carretero-Gonz{\'a}lez}}, \bibinfo {author} {\bibfnamefont {D.}~\bibnamefont
  {Khatri}}, \bibinfo {author} {\bibfnamefont {M.~A.}\ \bibnamefont {Porter}},
  \bibinfo {author} {\bibfnamefont {P.}~\bibnamefont {Kevrekidis}}, \ and\
  \bibinfo {author} {\bibfnamefont {C.}~\bibnamefont {Daraio}},\ }\href@noop {}
  {\bibfield  {journal} {\bibinfo  {journal} {Physical review letters}\
  }\textbf {\bibinfo {volume} {102}},\ \bibinfo {pages} {024102} (\bibinfo
  {year} {2009})}\BibitemShut {NoStop}%
\bibitem [{\citenamefont {Vergara}(2010)}]{vergara2010model}%
  \BibitemOpen
  \bibfield  {author} {\bibinfo {author} {\bibfnamefont {L.}~\bibnamefont
  {Vergara}},\ }\href@noop {} {\bibfield  {journal} {\bibinfo  {journal}
  {Physical review letters}\ }\textbf {\bibinfo {volume} {104}},\ \bibinfo
  {pages} {118001} (\bibinfo {year} {2010})}\BibitemShut {NoStop}%
\bibitem [{\citenamefont {James}(2021)}]{James_2021}%
  \BibitemOpen
  \bibfield  {author} {\bibinfo {author} {\bibfnamefont {G.}~\bibnamefont
  {James}},\ }\href {\doibase 10.1088/1361-6544/abdbbe} {\bibfield  {journal}
  {\bibinfo  {journal} {Nonlinearity}\ }\textbf {\bibinfo {volume} {34}},\
  \bibinfo {pages} {1758} (\bibinfo {year} {2021})}\BibitemShut {NoStop}%
\bibitem [{\citenamefont {Hoogeboom}\ \emph {et~al.}(2013)\citenamefont
  {Hoogeboom}, \citenamefont {Man}, \citenamefont {Boechler}, \citenamefont
  {Theocharis}, \citenamefont {Kevrekidis}, \citenamefont {Kevrekidis},\ and\
  \citenamefont {Daraio}}]{hoogeboom2013hysteresis}%
  \BibitemOpen
  \bibfield  {author} {\bibinfo {author} {\bibfnamefont {C.}~\bibnamefont
  {Hoogeboom}}, \bibinfo {author} {\bibfnamefont {Y.}~\bibnamefont {Man}},
  \bibinfo {author} {\bibfnamefont {N.}~\bibnamefont {Boechler}}, \bibinfo
  {author} {\bibfnamefont {G.}~\bibnamefont {Theocharis}}, \bibinfo {author}
  {\bibfnamefont {P.}~\bibnamefont {Kevrekidis}}, \bibinfo {author}
  {\bibfnamefont {I.}~\bibnamefont {Kevrekidis}}, \ and\ \bibinfo {author}
  {\bibfnamefont {C.}~\bibnamefont {Daraio}},\ }\href@noop {} {\bibfield
  {journal} {\bibinfo  {journal} {EPL (Europhysics Letters)}\ }\textbf
  {\bibinfo {volume} {101}},\ \bibinfo {pages} {44003} (\bibinfo {year}
  {2013})}\BibitemShut {NoStop}%
\bibitem [{\citenamefont {Charalampidis}\ \emph {et~al.}(2015)\citenamefont
  {Charalampidis}, \citenamefont {Li}, \citenamefont {Chong}, \citenamefont
  {Yang},\ and\ \citenamefont {Kevrekidis}}]{charalampidis2015time}%
  \BibitemOpen
  \bibfield  {author} {\bibinfo {author} {\bibfnamefont {E.~G.}\ \bibnamefont
  {Charalampidis}}, \bibinfo {author} {\bibfnamefont {F.}~\bibnamefont {Li}},
  \bibinfo {author} {\bibfnamefont {C.}~\bibnamefont {Chong}}, \bibinfo
  {author} {\bibfnamefont {J.}~\bibnamefont {Yang}}, \ and\ \bibinfo {author}
  {\bibfnamefont {P.~G.}\ \bibnamefont {Kevrekidis}},\ }\href@noop {}
  {\bibfield  {journal} {\bibinfo  {journal} {Mathematical Problems in
  Engineering}\ }\textbf {\bibinfo {volume} {2015}},\ \bibinfo {pages} {830978}
  (\bibinfo {year} {2015})}\BibitemShut {NoStop}%
\bibitem [{\citenamefont {Lydon}\ \emph {et~al.}(2015)\citenamefont {Lydon},
  \citenamefont {Theocharis},\ and\ \citenamefont
  {Daraio}}]{lydon2015nonlinear}%
  \BibitemOpen
  \bibfield  {author} {\bibinfo {author} {\bibfnamefont {J.}~\bibnamefont
  {Lydon}}, \bibinfo {author} {\bibfnamefont {G.}~\bibnamefont {Theocharis}}, \
  and\ \bibinfo {author} {\bibfnamefont {C.}~\bibnamefont {Daraio}},\
  }\href@noop {} {\bibfield  {journal} {\bibinfo  {journal} {Physical Review
  E}\ }\textbf {\bibinfo {volume} {91}},\ \bibinfo {pages} {023208} (\bibinfo
  {year} {2015})}\BibitemShut {NoStop}%
\bibitem [{\citenamefont {Lee}\ \emph {et~al.}(2023)\citenamefont {Lee},
  \citenamefont {Charalampidis}, \citenamefont {Xing}, \citenamefont {Chong},\
  and\ \citenamefont {Kevrekidis}}]{pdimer}%
  \BibitemOpen
  \bibfield  {author} {\bibinfo {author} {\bibfnamefont {M.~M.}\ \bibnamefont
  {Lee}}, \bibinfo {author} {\bibfnamefont {E.~G.}\ \bibnamefont
  {Charalampidis}}, \bibinfo {author} {\bibfnamefont {S.}~\bibnamefont {Xing}},
  \bibinfo {author} {\bibfnamefont {C.}~\bibnamefont {Chong}}, \ and\ \bibinfo
  {author} {\bibfnamefont {P.~G.}\ \bibnamefont {Kevrekidis}},\ }\href
  {\doibase 10.1103/PhysRevE.107.054208} {\bibfield  {journal} {\bibinfo
  {journal} {Phys. Rev. E}\ }\textbf {\bibinfo {volume} {107}},\ \bibinfo
  {pages} {054208} (\bibinfo {year} {2023})}\BibitemShut {NoStop}%
\bibitem [{\citenamefont {Doedel}\ \emph {et~al.}(1991)\citenamefont {Doedel},
  \citenamefont {Keller},\ and\ \citenamefont
  {Kernevez}}]{doedel1991numerical}%
  \BibitemOpen
  \bibfield  {author} {\bibinfo {author} {\bibfnamefont {E.}~\bibnamefont
  {Doedel}}, \bibinfo {author} {\bibfnamefont {H.~B.}\ \bibnamefont {Keller}},
  \ and\ \bibinfo {author} {\bibfnamefont {J.~P.}\ \bibnamefont {Kernevez}},\
  }\href@noop {} {\bibfield  {journal} {\bibinfo  {journal} {International
  journal of bifurcation and chaos}\ }\textbf {\bibinfo {volume} {1}},\
  \bibinfo {pages} {493} (\bibinfo {year} {1991})}\BibitemShut {NoStop}%
\bibitem [{\citenamefont {Guckenheimer}\ and\ \citenamefont
  {Holmes}(1983)}]{guckenheimer1983nonlinear}%
  \BibitemOpen
  \bibfield  {author} {\bibinfo {author} {\bibfnamefont {J.}~\bibnamefont
  {Guckenheimer}}\ and\ \bibinfo {author} {\bibfnamefont {P.}~\bibnamefont
  {Holmes}},\ }\href@noop {} {\emph {\bibinfo {title} {Nonlinear oscillations,
  dynamical systems, and bifurcations of vector fields}}}\ (\bibinfo
  {publisher} {Springer},\ \bibinfo {year} {1983})\BibitemShut {NoStop}%
\bibitem [{\citenamefont {Kevrekidis}(1987)}]{kevrekidis1987numerical}%
  \BibitemOpen
  \bibfield  {author} {\bibinfo {author} {\bibfnamefont {I.}~\bibnamefont
  {Kevrekidis}},\ }\href@noop {} {\bibfield  {journal} {\bibinfo  {journal}
  {AIChE journal}\ }\textbf {\bibinfo {volume} {33}},\ \bibinfo {pages} {1850}
  (\bibinfo {year} {1987})}\BibitemShut {NoStop}%
\bibitem [{\citenamefont {Hobson}(1993)}]{hobson1993efficient}%
  \BibitemOpen
  \bibfield  {author} {\bibinfo {author} {\bibfnamefont {D.}~\bibnamefont
  {Hobson}},\ }\href@noop {} {\bibfield  {journal} {\bibinfo  {journal}
  {Journal of Computational Physics}\ }\textbf {\bibinfo {volume} {104}},\
  \bibinfo {pages} {14} (\bibinfo {year} {1993})}\BibitemShut {NoStop}%
\bibitem [{\citenamefont {Krauskopf}\ and\ \citenamefont
  {Osinga}(1998)}]{krauskopf1998growing}%
  \BibitemOpen
  \bibfield  {author} {\bibinfo {author} {\bibfnamefont {B.}~\bibnamefont
  {Krauskopf}}\ and\ \bibinfo {author} {\bibfnamefont {H.}~\bibnamefont
  {Osinga}},\ }\href@noop {} {\bibfield  {journal} {\bibinfo  {journal}
  {Journal of Computational Physics}\ }\textbf {\bibinfo {volume} {146}},\
  \bibinfo {pages} {404} (\bibinfo {year} {1998})}\BibitemShut {NoStop}%
\bibitem [{\citenamefont {Kaneko}(1983)}]{kaneko1983doubling}%
  \BibitemOpen
  \bibfield  {author} {\bibinfo {author} {\bibfnamefont {K.}~\bibnamefont
  {Kaneko}},\ }\href@noop {} {\bibfield  {journal} {\bibinfo  {journal}
  {Progress of theoretical physics}\ }\textbf {\bibinfo {volume} {69}},\
  \bibinfo {pages} {1806} (\bibinfo {year} {1983})}\BibitemShut {NoStop}%
\bibitem [{\citenamefont {Franceschini}(1983)}]{franceschini1983bifurcations}%
  \BibitemOpen
  \bibfield  {author} {\bibinfo {author} {\bibfnamefont {V.}~\bibnamefont
  {Franceschini}},\ }\href@noop {} {\bibfield  {journal} {\bibinfo  {journal}
  {Physica D: Nonlinear Phenomena}\ }\textbf {\bibinfo {volume} {6}},\ \bibinfo
  {pages} {285} (\bibinfo {year} {1983})}\BibitemShut {NoStop}%
\bibitem [{\citenamefont {Iooss}\ and\ \citenamefont
  {Los}(1988)}]{iooss1988quasi}%
  \BibitemOpen
  \bibfield  {author} {\bibinfo {author} {\bibfnamefont {G.}~\bibnamefont
  {Iooss}}\ and\ \bibinfo {author} {\bibfnamefont {J.}~\bibnamefont {Los}},\
  }\href@noop {} {\bibfield  {journal} {\bibinfo  {journal} {Communications in
  Mathematical Physics}\ }\textbf {\bibinfo {volume} {119}},\ \bibinfo {pages}
  {453} (\bibinfo {year} {1988})}\BibitemShut {NoStop}%
\bibitem [{\citenamefont {Grebogi}\ \emph
  {et~al.}(1987{\natexlab{a}})\citenamefont {Grebogi}, \citenamefont {Ott},\
  and\ \citenamefont {Yorke}}]{A1}%
  \BibitemOpen
  \bibfield  {author} {\bibinfo {author} {\bibfnamefont {C.}~\bibnamefont
  {Grebogi}}, \bibinfo {author} {\bibfnamefont {E.}~\bibnamefont {Ott}}, \ and\
  \bibinfo {author} {\bibfnamefont {J.~A.}\ \bibnamefont {Yorke}},\ }\href
  {http://www.jstor.org/stable/1700479} {\bibfield  {journal} {\bibinfo
  {journal} {Science}\ }\textbf {\bibinfo {volume} {238}},\ \bibinfo {pages}
  {632} (\bibinfo {year} {1987}{\natexlab{a}})}\BibitemShut {NoStop}%
\bibitem [{\citenamefont {Grebogi}\ \emph
  {et~al.}(1987{\natexlab{b}})\citenamefont {Grebogi}, \citenamefont {Ott},
  \citenamefont {Romeiras},\ and\ \citenamefont {Yorke}}]{A2}%
  \BibitemOpen
  \bibfield  {author} {\bibinfo {author} {\bibfnamefont {C.}~\bibnamefont
  {Grebogi}}, \bibinfo {author} {\bibfnamefont {E.}~\bibnamefont {Ott}},
  \bibinfo {author} {\bibfnamefont {F.}~\bibnamefont {Romeiras}}, \ and\
  \bibinfo {author} {\bibfnamefont {J.~A.}\ \bibnamefont {Yorke}},\ }\href
  {\doibase 10.1103/PhysRevA.36.5365} {\bibfield  {journal} {\bibinfo
  {journal} {Phys. Rev. A}\ }\textbf {\bibinfo {volume} {36}},\ \bibinfo
  {pages} {5365} (\bibinfo {year} {1987}{\natexlab{b}})}\BibitemShut {NoStop}%
\bibitem [{\citenamefont {Johnson}\ \emph {et~al.}(2001)\citenamefont
  {Johnson}, \citenamefont {Jolly},\ and\ \citenamefont {Kevrekidis}}]{A3}%
  \BibitemOpen
  \bibfield  {author} {\bibinfo {author} {\bibfnamefont {M.~E.}\ \bibnamefont
  {Johnson}}, \bibinfo {author} {\bibfnamefont {M.~S.}\ \bibnamefont {Jolly}},
  \ and\ \bibinfo {author} {\bibfnamefont {I.~G.}\ \bibnamefont {Kevrekidis}},\
  }\href {https://api.semanticscholar.org/CorpusID:29854929} {\bibfield
  {journal} {\bibinfo  {journal} {Int. J. Bifurc. Chaos}\ }\textbf {\bibinfo
  {volume} {11}},\ \bibinfo {pages} {1} (\bibinfo {year} {2001})}\BibitemShut
  {NoStop}%
\end{thebibliography}%

\end{document}